\documentclass[aps,prd,twocolumn,groupedaddress,showkeys,floatfix]{revtex4-1} 

\usepackage{graphicx}
\usepackage{amssymb}
\usepackage{amsmath}
\usepackage{bm}
\newcommand{\op}[1]{\mathrm{\bf #1}}

\begin{document}


\title{A compendium of photon emission rates, 
absorption cross sections and scattering cross sections}




\author{Rainer Dick}
\email{rainer.dick@usask.ca}
\affiliation{Department of Physics and Engineering Physics, 
University of Saskatchewan, 116 Science Place, Saskatoon, Canada SK S7N 5E2}



\begin{abstract} 
We provide a compendium of the quantum mechanical equations for
photon emission rates, photon absorption cross sections, and photon scattering cross sections.
 For each case, the different equations that apply for discrete or continuous electron states
of the emitting, absorbing, or scattering material are given.
\end{abstract}


\maketitle



\section{Introduction\label{sec:intro}}

Photons can be emitted, absorbed, or scattered in materials. Furthermore, the initial and final 
quantum states in atoms, molecules or condensed materials can be discrete or continuous. This 
allows for twelve different kinds of basic quantum optical transitions, and each of these kinds 
of transitions are described by their own quantum mechanical formulae.

The relevant techniques to
derive the formulae involve second quantization and time-dependent perturbation theory, which are 
described in textbooks on advanced quantum mechanics, 
see e.g.~\cite{Messiah,Sakurai,Merzbacher,Schwabl,Dick}. 
These techniques are applied to the
quantum optics Hamiltonian to derive amplitudes for photon emission, absorption and scattering.
However, our focus in the present paper is not a review of the pertinent theoretical techniques, 
but rather to provide a concise list
of the resulting photon emission rates, absorption cross sections, and scattering cross sections
in the cases where photon-matter interactions are dominated by 
minimal coupling between photons and nonrelativistic electrons.
We will therefore only very briefly review the quantum optics
Hamiltonian in Eqs.~(\ref{eq:HQO}-\ref{eq:Hegamma}).
Readers who are not interested in these details may very well skip
Eqs.~(\ref{eq:HQO}-\ref{eq:HQOab})
and only take note that the results collected in this compendium are
derived from the standard electron-photon coupling terms (\ref{eq:Hegamma}).

The description of photon processes in many-particle systems
requires the use of quantized electromagnetic potentials for the photons and
quantized matter fields $\Psi(\bm{x},t)$ for the charged particles,
and therefore a field-theoretic formulation of the Hamiltonian.
The quantum optics Hamiltonian in Coulomb gauge, $\bm{\nabla}\cdot\bm{A}(\bm{x},t)=0$,
takes the form \cite{Dick}
\begin{eqnarray}\nonumber
H&=&\int\!d^3\bm{x}\left(\sum_a\mathcal{H}_a(\bm{x},t)+\mathcal{H}_\gamma(\bm{x},t)
+\sum_a\mathcal{H}_{a\gamma}(\bm{x},t)\right)
\\ \label{eq:HQO}
&&+\int\!d^3\bm{x}\int\!d^3\bm{x}'\sum_{ab}\mathcal{H}_{ab}(\bm{x},\bm{x}',t).
\end{eqnarray}
Here
\begin{equation}\label{eq:HQOa}
\mathcal{H}_a(\bm{x},t)=\sum_s\frac{\hbar^2}{2m_a}
\bm{\nabla}\Psi_{a,s}^+(\bm{x},t)\cdot\bm{\nabla}\Psi_{a,s}(\bm{x},t)
\end{equation}
is the kinetic term for particles of species $a$ (electrons and atomic nuclei)
with mass $m_a$ and electric charge $q_a$, and $s$ is a spin orientation label.
 Furthermore,
\begin{equation}\label{eq:HQOgamma}
\mathcal{H}_\gamma(\bm{x},t)=\frac{\epsilon_0}{2}
\left(\frac{\partial\bm{A}(\bm{x},t)}{\partial t}\right)^2
+\frac{1}{2\mu_0}\bm{B}^2(\bm{x},t)
\end{equation}
is the kinetic photon term,
\begin{eqnarray}\nonumber
\mathcal{H}_{a\gamma}(\bm{x},t)&=&\frac{1}{2m_a}\sum_s\left[
\mathrm{i}q_a\hbar\Psi^+_{a,s}(\bm{x},t)\bm{A}(\bm{x},t)\cdot\bm{\nabla}\Psi_{a,s}(\bm{x},t)
\right.
\\ \nonumber
&&-\,
\mathrm{i}q_a\hbar\bm{\nabla}\Psi^+_{a,s}(\bm{x},t)\cdot\bm{A}(\bm{x},t)\Psi_{a,s}(\bm{x},t)
\\ \label{eq:Hagamma}
&&+\left. q_a^2\Psi_{a,s}^+(\bm{x},t)\bm{A}^2(\bm{x},t)
\Psi_{a,s}(\bm{x},t)\right]
\end{eqnarray}
is the matter-photon interaction term for particles of species $a$,
and 
\begin{eqnarray}\nonumber
\mathcal{H}_{ab}(\bm{x},\bm{x}',t)&=&\sum_{s,s'}\frac{q_aq_{b}}{8\pi\epsilon_0|\bm{x}-\bm{x}'|}
\Psi_{a,s}^+(\bm{x},t)\Psi_{b,s'}^+(\bm{x}',t)
\\ \label{eq:HQOab}
&&\times
\Psi_{b,s'}(\bm{x}',t)\Psi_{a,s}(\bm{x},t)
\end{eqnarray}
is the Coulomb interaction term between particle species $a$ and $b$.
 All operator products are assumed to be normal ordered.

The mass dependence of $\mathcal{H}_{a\gamma}$ implies that matter-photon interactions are
generically dominated by the electron-photon interaction terms 
(in the following we omit the electron labels: $m_e=m$, $\Psi_{e,s}(\bm{x},t)=\Psi_{s}(\bm{x},t)$),
\begin{eqnarray}\nonumber
\mathcal{H}_{e\gamma}(\bm{x},t)&=&\frac{1}{2m}\sum_s\left[
\mathrm{i}e
\hbar\bm{\nabla}\Psi^+_{s}(\bm{x},t)\cdot\bm{A}(\bm{x},t)\Psi_{s}(\bm{x},t)
\right.
\\ \nonumber
&&-\,\mathrm{i}e\hbar\Psi^+_{s}(\bm{x},t)\bm{A}(\bm{x},t)\cdot\bm{\nabla}\Psi_{s}(\bm{x},t)
\\ \label{eq:Hegamma}
&&+\left. e^2\Psi_{s}^+(\bm{x},t)\bm{A}^2(\bm{x},t)
\Psi_{s}(\bm{x},t)\right].
\end{eqnarray}

The quantization conditions on the electron fields $\Psi_{s}(\bm{x},t)$
and photon fields $\bm{A}(\bm{x},t)$, and the relation between the electron field 
operators and electronic states and wave functions are briefly reviewed in Appendix B.

The minimal coupling terms (\ref{eq:Hegamma}) dominate photon-matter interactions
in the visible, UV and X-ray regime, and depending on the material under study,
they can also dominate in the infrared regime. 
Exceptions can occur in the infrared
wavelength range due to molecular vibrations and rotations, and in the microwave regime
where spin-flipping Pauli terms can dominate due to the unavailability of
spin-preserving electronic transitions.

Practitioners of photonics or spectroscopy are generically not interested in the
technical details of the derivations of the 
pertinent quantum mechanical formulae for each particular kind of
transition, while on the other hand there is no concise 
overview available of the pertinent formulae that covers all twelve types of optical transitions. 
The purpose of the present paper is therefore to provide such an overview as a resource for easy 
reference and for easy comparison of the formulae that apply to the different situations.

Indeed, the corresponding photon emission or absorption rates for transitions between discrete 
electronic energy levels are standard textbook examples, and photon absorption rates due to ionization
are often reported through the Golden Rule. Furthermore, numerous applications of photon emission
rates and absorption cross sections can be found in atomic, molecular and optical physics and
throughout the spectroscopic literature, see
e.g.~\cite{Drake} and references there. The unpolarized equations for photon absorption and
emission due to transitions between discrete
electronic states have also been compiled already by
Hilborn \cite{Hilborn}, and
the photon scattering cross section (including the Thomson term)
due to transitions between discrete electron states is
reported e.g.~in \cite{Sakurai,Dick}.
However, the corresponding equations for interband or intraband transitions in materials
have never been discussed in a concise review, nor have all twelve cases
of optical transitions been summarized in a single reference before. 

Expressions in terms of densities of states in the energy scale are popular to describe transition
rates and cross sections involving continuous electronic states in the material. However, this requires
labeling of the continuous states in the form $|E,\nu\rangle$ with discrete degeneracy indices $\nu$.
This is in principle always possible through harmonic analysis, but 
is practical only if the particles are moving in a radially symmetric potential, when angular momentum 
quantum numbers $\nu=\{\ell,m_\ell\}$ provide discrete degeneracy indices on the constant energy 
surfaces $E(k_e)=E$. Formulations in terms of densities of states in the energy scale are therefore 
useful for the calculation of ionization rates in atomic physics.
However, another practically important case of optical transitions involving continuous electron states 
concerns transitions from or into energy bands in materials, and in these
cases we should formulate transition rates and cross sections in terms of Bloch energy
eigenstates $|n,\bm{k}_e\rangle$. We will therefore provide equations in both formalisms.

All equations are given in dipole approximation, which is suitable up into the soft
X-ray regime $E_\gamma\lesssim 1\,\mathrm{keV}$, $\lambda\gtrsim 1\,\mathrm{nm}$, 
and most results are displayed in ``length form'', 
i.e.~in terms of matrix elements of the position operator $\op{x}$,
\begin{equation}\label{eq:lengthform}
\langle f|\op{x}|i\rangle=\int\!d^3\bm{x}\,\psi^+_f(\bm{x})\bm{x}\psi_i(\bm{x}).
\end{equation}
Here we use upright notation for the quantum mechanical position operator $\op{x}$
or the momentum operator $\op{p}$ and standard italics notation for the corresponding classical 
vectors $\bm{x}$ or $\bm{p}$. We also use upright notation for $\mathrm{i}=\sqrt{-1}$ to
distinguish it from labels for initial states $|i\rangle$.
The initial and final wave functions $\psi_i(\bm{x})$ and $\psi_f(\bm{x})$
refer to energy eigenfunctions with corresponding
eigenvalues $E_i$ and $E_f$ of $H$. At the level of the spectroscopic formulae compiled in this
overview, the wave functions and energy eigenvalues refer to single nonrelativistic electrons moving
in a potential $V(\op{x})$, such that the first-quantized Hamiltonian is
\begin{equation}\label{eq:Hfirstq}
H=\frac{\op{p}^2}{2m}+V(\op{x}).
\end{equation}
The mass $m\equiv m_e=511$ keV is the electron mass, because in many-electron atoms, in molecules,
or in solid materials,
the photons will couple to fundamental electrons through minimal coupling \cite{notequasi1}.
The emergence of effective single-electron Hamiltonians of the form (\ref{eq:Hfirstq})
in many-electron systems from the quantum optics Hamiltonian (\ref{eq:HQO}) is briefly outlined
in Appendix B.

Translation from the length form (\ref{eq:lengthform}) into the ``velocity form'' in terms of 
matrix elements $\langle f|\op{p}|i\rangle$ of the momentum operator $\op{p}$ proceeds through
\begin{equation}\label{eq:l2v}
\langle f|\op{x}|i\rangle=\frac{\langle f|\op{p}|i\rangle}{\mathrm{i}m\omega_{fi}},
\end{equation}
with the transition frequency
\begin{equation}
\omega_{fi}=-\,\omega_{if}=\frac{E_f-E_i}{\hbar}.
\end{equation}
The transition rates in terms of the length form or the equivalent velocity form 
use the assumption that the electron states involved with photon emission, absorption
or scattering 
are described through eigenstates of Hamiltonians of the form (\ref{eq:Hfirstq}).
The electron-photon coupling term (\ref{eq:Hegamma})
implies $\op{p}\to\op{p}+e\op{A}(\op{x},t)$ in (\ref{eq:Hfirstq})
and yields the leading-order
electron-photon coupling terms $e[\op{p}\cdot\op{A}(\op{x},t)+\op{A}(\op{x},t)\cdot\op{p}]/2m$.
This yields transition rates in velocity form in the first place. 
However, use of the length form is much more common and therefore this convention is also 
adopted here. All the length-form results listed below can be transformed back 
into the corresponding velocity forms through the substitution (\ref{eq:l2v}).

To keep the presentation as concise and useful as possible, technical remarks are 
kept to a minimum. However, details from the derivations of the formulae are included if they help
to understand the structure and physical interpretations of the formulae. We denote initial 
and final electron states with $|i\rangle$ and $|f\rangle$, respectively. The
letters $i$ 
and $f$ serve as placeholders for complete sets of quantum numbers for the electron states.  We also 
indicate the presence of a photon with $\gamma$, e.g. $|f,\gamma\rangle$ if there is also a photon 
in the final state. The letter $\gamma$ serves as a placeholder for photon momentum $\hbar{\bm{k}}$ 
and polarization vector $\bm{\epsilon}_\gamma$.  

The formulae for photon emission are listed in Sec.~\ref{sec:em}, for photon absorption in 
Sec.~\ref{sec:abs}, and for photon scattering in Sec.~\ref{sec:sc}. Within each Section, the
different cases for the transitions $|i\rangle\to|f\rangle$ corresponding to discrete $\to$ discrete, 
discrete $\to$ continuous, continuous $\to$ discrete and continuous $\to$ continuous are organized in 
subsections, thus generating a catalogue and quasi-tabular overview of the pertinent formulae. Brief
discussions of common labelings for continuous states, and of the quantum fields of quantum optics,
are provided in Appendices A and B, respectively.
Recoil effects are briefly discussed in Appendix C, and Appendix
D contains a brief discussion of radiative electron-hole recombination in materials.

In keeping with the user-oriented spirit of this paper, a few formulations will be redundant
between different subsections to ensure that a reader who is primarily interested e.g.~in
emission from decay of an acceptor state (i.e.~emission due to a discrete $\to$ continuous
electronic transition) can directly jump to subsection \ref{sub:emdc} for the basic emission
rates without having to consult the other subsections.

With respect to some special notations used in this paper, $V$ without any argument denotes
the volume of the Wigner-Seitz cell in a lattice, whereas the function $V(\bm{x})$ denotes a
potential. The symbol $\mathcal{V}$ without argument
denotes the spatial volume factor in Fermi's trick
$\delta^2(\bm{k})\to \delta(\bm{k})\mathcal{V}/8\pi^3$ which also
appears in the corresponding elementary
volume unit in $\bm{k}$-space, $\Delta^3\bm{k}=8\pi^3/\mathcal{V}$.
In a crystal, we express this volume also as a sum of $N_{\bm{R}}$
Wigner-Seitz cells, $\mathcal{V}=N_{\bm{R}}V$.
In Appendix B, the symbol $\mathcal{V}_e(\bm{x})$ denotes a potential operator
in terms of quantized fields.

We use $\varrho_V(E)$ as the density
of quantum states in the energy scale (e.g.~in units of $\mathrm{eV}^{-1}$)
in a volume $V$, such that
\begin{equation}
N_{[E_1,E_2]}(V)=\int_{E_1}^{E_2}\!dE\,\varrho_V(E)
\end{equation}
is the number of quantum states with energies in the range $E_1\le E\le E_2$
in the volume $V$.
The density of states $\varrho_V(E)$ is related to the local density of
states $\varrho(E,\bm{x})$ (e.g.~in units of $\mathrm{eV}^{-1}\mathrm{nm}^{-3}$)
through volume integration,
\begin{equation}
\varrho_V(E)=\int_V\!d^3\bm{x}\,\varrho(E,\bm{x}).
\end{equation}
Another manifestation $\tilde{\varrho}(E,\nu)$ of the density of states arises from
contributions of the continuous parts $C$ of the spectrum to completeness relations,
\begin{eqnarray}\nonumber
1&=&\sum_{j,\nu}|E_j,\nu\rangle\langle E_j,\nu|
\\ \label{eq:completeE1}
&&+\sum_{\nu}\int_{C}\!dE\,|E,\nu\rangle\tilde{\varrho}(E,\nu)\langle E,\nu|.
\end{eqnarray}
Here $E_j$ enumerates the discrete energy eigenvalues and $\nu$ is the set
of degeneracy indices. The normalization and dimensions
of the measure factors $\tilde{\varrho}(E,\nu)$ depend on the normalization and
dimensions of the continuous energy eigenstates $|E,\nu\rangle$.
The local density of states is given by the measure factors $\tilde{\varrho}(E,\nu)$
and the energy eigenfunctions through 
\begin{eqnarray}\nonumber
  \varrho(E,\bm{x})&=&\sum_{j,\nu}|\langle\bm{x}|E_j,\nu\rangle|^2\delta(E-E_j)
  \\ \label{eq:varrhoEx}
  &&+\sum_{\nu}|\langle\bm{x}|E,\nu\rangle|^2\tilde{\varrho}(E,\nu),
\end{eqnarray}
see Appendix A for examples. Densities of states $\tilde{\varrho}(E,\nu)$
appear in transition rates involving electron states $|E,\nu\rangle$
in the continuous part of the spectrum.
Densities of states $\varrho_V(E)$ occur in equations for transition rates 
involving initial or final states in an energy band, and $V$ is
the volume of the Wigner-Seitz cell in these cases.

\section{Photon emission $\bm{|i\rangle\to |f,\gamma\rangle}$}
\label{sec:em}

Besides momentum $\hbar\bm{k}$, the
polarization $\bm{\epsilon}$ is another basic property of photons.
Polarization corresponds to a normalized vector, $\bm{\epsilon}^2=1$,
 that is perpendicular to the photon wave vector, $\bm{\epsilon}\cdot\bm{k}=0$.
As such, polarization can be expressed for every wave vector $\bm{k}$ as a linear combination
of a two-dimensional orthonormal basis that spans the plane perpendicular to $\bm{k}$,
$\bm{\epsilon}=\sum_\gamma c_\gamma\bm{\epsilon}_{\gamma}(\bm{k})$, $\sum_\gamma |c_\gamma|^2=1$.
Both Cartesian bases $\{\bm{\epsilon}_{1}(\bm{k}),\bm{\epsilon}_{2}(\bm{k})\}$
and circularly polarized bases with vectors 
$\bm{\epsilon}_{\pm}(\bm{k})=[\bm{\epsilon}_{1}(\bm{k})\pm\mathrm{i}\bm{\epsilon}_{2}(\bm{k})]/\sqrt{2}$ 
are commonly used.
Either way, summation over the tensor products of the basis vectors generates a $3\times 3$
matrix that projects every vector onto the plane orthogonal to the wave vector,
\begin{equation}
\sum_\gamma\bm{\epsilon}_{\gamma}(\bm{k})\otimes\bm{\epsilon}^+_{\gamma}(\bm{k})
=\underline{1}-\hat{\bm{k}}\otimes\hat{\bm{k}}^T.
\end{equation}
Here $\hat{\bm{k}}=\bm{k}/|\bm{k}|$ is the unit wave vector in the direction of photon motion.
We will denote the polarization vector of a photon 
with $\bm{\epsilon}_{\gamma}(\bm{k})\equiv\bm{\epsilon}_{\gamma}$
in the following, i.e.~without the explicit reminder that it depends on $\bm{k}$
through the requirement of orthogonality. 

Depending on instrumentation and available photon beams, the practically relevant observables for 
photon emission concern polarized differential emission rates $d\Gamma_{fi}/d\Omega$, unpolarized
differential emission rates $d\tilde{\Gamma}_{fi}/d\Omega=\sum_{\bm{\epsilon}_{\gamma}}d\Gamma_{fi}/d\Omega$, 
and unpolarized total emission rates
$\tilde{\Gamma}_{fi}=\int\!d\Omega\,d\tilde{\Gamma}_{fi}/d\Omega$ \cite{Note1}.
The basic equations for all these quantities follow simple translation rules in terms
of substitutions and scalings of the basic dipole
transition factor $|\langle f|\bm{\epsilon}^+_{\gamma}\cdot\op{x}|i\rangle|^2$:

Differential emission rates $d\Gamma_{fi}/d\Omega$ for photons with 
polarization $\bm{\epsilon}_{\gamma}$ depend on
 $|\langle f|\bm{\epsilon}^+_{\gamma}\cdot\op{x}|i\rangle|^2$. For the translation 
into \textit{unpolarized} differential emission rates $d\tilde{\Gamma}_{fi}/d\Omega$
we note that summation over photon 
polarizations for photons with momentum $\hbar{\bm{k}}$ yields
\begin{equation}\label{eq:unpol}
\sum_{\bm{\epsilon}_\gamma}|\langle f|\bm{\epsilon}^+_\gamma\cdot\op{x}|i\rangle|^2
=|\langle f|\op{x}|i\rangle|^2
-|\langle f|\hat{\bm{k}}\cdot\op{x}|i\rangle|^2.
\end{equation}

For the translation of the polarized differential emission rate $d\Gamma_{fi}/d\Omega$
into the 
polarized total emission rate $\Gamma_{fi}$ we note
\begin{equation}\label{eq:total1}
\int\!d\Omega\,|\langle f|\bm{\epsilon}^+_{\gamma}\cdot\op{x}|i\rangle|^2=\frac{4\pi}{3}
|\langle f|\op{x}|i\rangle|^2.
\end{equation}
The integration over angles has removed the dependence on the polarization.
The equations for total unpolarized photon emission 
rates $\tilde{\Gamma}_{fi}=\sum_{\bm{\epsilon}_{\gamma}}\int\!d\Omega\,d\Gamma_{fi}/d\Omega$
therefore satisfy $\tilde{\Gamma}_{fi}=2\Gamma_{fi}$
and follow from the equations for $d\Gamma_{fi}/d\Omega$ through the substitution
\begin{equation}\label{eq:total2}
  |\langle f|\bm{\epsilon}^+_{\gamma}\cdot\op{x}|i\rangle|^2
  \to\frac{8\pi}{3}|\langle f|\op{x}|i\rangle|^2.
\end{equation}
Eqs.~(\ref{eq:emdd1},\ref{eq:emdd2}) provide an example for the substitution rule.

Eq.~(\ref{eq:total1}) is a consequence of the fact that for any two real
vectors $\bm{\epsilon}$ and $\bm{d}$ the equations
\begin{equation}\label{eq:avproof1}
\int\!d\Omega\,(\bm{\epsilon}\cdot\bm{d})^2=\frac{4\pi}{3}\bm{\epsilon}^2\bm{d}^2
\end{equation}
and
\begin{equation}\label{eq:avproof2}
\int\!d\Omega\,\bm{\epsilon}\cdot\bm{d}=0
\end{equation}
hold.
The vector
\begin{equation}
  \langle f|\op{x}|i\rangle=\bm{d}_{fi}^{(1)}
  +\mathrm{i}\bm{d}_{fi}^{(2)}
\end{equation}
will generically be complex, and the polarization vector
\begin{equation}
\bm{\epsilon}_{\gamma}=\bm{\epsilon}_{\gamma}^{(1)}
+\mathrm{i}\bm{\epsilon}_{\gamma}^{(2)},\quad
(\bm{\epsilon}_{\gamma}^{(1)})^2+(\bm{\epsilon}_{\gamma}^{(2)})^2=1,
\end{equation}
can be complex if we choose a chiral polarization basis.
However, we can express the integrand in (\ref{eq:total1})
in terms of products of real vectors in the form
\begin{eqnarray}\nonumber
  |\langle f|\bm{\epsilon}^+_{\gamma}\cdot\op{x}|i\rangle|^2
  &=&2(\bm{\epsilon}_{\gamma}^{(1)}\times\bm{\epsilon}_{\gamma}^{(2)})
  \cdot(\bm{d}_{fi}^{(1)}\times\bm{d}_{fi}^{(2)})
  \\ \label{eq:avproof4}
  &&+\,\sum_{ij}(\bm{\epsilon}_{\gamma}^{(i)}\cdot\bm{d}_{fi}^{(j)})^2,
\end{eqnarray}
and this shows that Eq.~(\ref{eq:total1}) arises as a consequence of
Eqs.~(\ref{eq:avproof1},\ref{eq:avproof2}).

\subsection{Emission rates for both $\bm{|i\rangle}$ and $\bm{|f\rangle}$ discrete}
\label{sub:emdd}

This case applies e.g.~to spontaneous photon emission due to transitions
$|i\rangle\to|f\rangle$, $E_i>E_f$, between bound states in atoms or molecules.
The differential emission rate for photons with polarization $\bm{\epsilon}_\gamma$
into a solid angle $d\Omega$ in the direction $\hat{\bm{k}}$ is
\begin{equation}\label{eq:emdd1}
\frac{d\Gamma_{fi}}{d\Omega}=\frac{\alpha_S}{2\pi c^2}\omega_{if}^3
|\langle f|\bm{\epsilon}^+_\gamma\cdot\op{x}|i\rangle|^2,
\end{equation}
where $\alpha_S=e^2/4\pi\epsilon_0\hbar c$ is the fine structure constant.
The total unpolarized emission rate is the Einstein $A$ coefficient for
the transition $|i\rangle\to|f\rangle$,
\begin{equation}\label{eq:emdd2}
A_{fi}\equiv\tilde{\Gamma}_{fi}=\frac{4\alpha_S}{3c^2}\omega_{if}^3
|\langle f|\op{x}|i\rangle|^2,
\end{equation}
see also \cite{Hilborn}, where the relations to Einstein's $B$ coefficients
are also reviewed.

The initial and final states will initially depend on the spin projections
(magnetic quantum numbers) $s_i$ and $s_f$ of the electrons in the initial
state $|i\rangle$ and the final state $|f\rangle$, respectively. If we are
not explicitly interested in transitions between spin-polarized states,
transition rates are calculated in general through averaging
over initial spin projection and summation over final spin projection, e.g.
\begin{equation}\label{eq:average1}
  \frac{d\Gamma_{fi}}{d\Omega}=\frac{1}{2}\sum_{s_f=-1/2}^{1/2}\sum_{s_i=-1/2}^{1/2}
  \frac{d\Gamma_{fi}}{d\Omega}(s_f,s_i).
\end{equation}
The leading order electron-photon couplings (\ref{eq:Hegamma}) used in this
review do not induce spin flips and Eq.~(\ref{eq:average1})
reduces to
\begin{equation}\label{eq:average2}
  \frac{d\Gamma_{fi}}{d\Omega}=\frac{1}{2}\sum_{s=-1/2}^{1/2}
  \frac{d\Gamma_{fi}}{d\Omega}(s,s)
\end{equation}
for transitions through the interaction terms in Eq.~(\ref{eq:Hegamma}).
Note that the terms \textit{polarized} or \textit{unpolarized} in this
paper always refer to photon polarization. Otherwise, we will use the term
\textit{spin-polarized}. Transition rates can depend on spin orientation if
there are spin-polarized energy bands in materials as a consequence of spin-orbit
coupling, exchange interactions, or external magnetic fields.
All the formulae reported in this paper apply to these
cases if spin orientation is properly included with the quantum numbers of states,
and if Eq.~(\ref{eq:average2}) is used to derive spin-averaged transition rates.

Unpolarized emission rates like $\tilde{\Gamma}_{fi}$ or the corresponding absorption
rates are traditionally often expressed in terms of the 
oscillator strengths,
\begin{equation}\label{eq:Ffi}
  F_{fi}=\frac{2m}{3\hbar}\omega_{fi}|\langle f|\op{x}|i\rangle|^2
  =-\,\frac{2m}{3\hbar}\omega_{if}|\langle f|\op{x}|i\rangle|^2,
\end{equation}
see e.g.~\cite{Merzbacher} (where the corresponding one-dimensional definition is
introduced) or \cite{Dick}.
Here we use the less common designation of capital $F$ for the oscillator strength
to avoid confusion with the label $f$ for the final state.

The oscillator strengths (\ref{eq:Ffi}) are dimensionless if both 
electronic states are discrete.
Otherwise, they have dimensions commensurate
with the labeling of the continuous states and their contributions to
the completeness relations (\ref{eq:completeE},\ref{eq:completeu}).
A nice feature of the oscillator strengths for
discrete initial states $|i\rangle$ is the existence of sum rules when
summed over the final states $|f\rangle$, where the 
appropriate measure factors from (\ref{eq:completeE},\ref{eq:completeu})
need to be included when summing over the continuous parts of the spectrum.
We will report all emission rates, absorption cross sections,
and scattering cross sections directly in terms of the dipole matrix elements
 for easier comparison of the cases with discrete and continuous initial states.

Atomic recoils do not change
the forms of Eqs.~(\ref{eq:emdd1},\ref{eq:emdd2}) because they do not introduce
additional factors in the final results for emission rates. They only shift the
transition frequency $\omega_{if}$ by small amounts, see Appendix C and in
particular Eq.~(\ref{eq:DeltaE}).

\subsection{Emission rates for $\bm{|i\rangle}$ discrete and $\bm{|f\rangle}$ continuous}
\label{sub:emdc}

This situation applies e.g.~to electron-hole recombination 
if the electron was stored in an acceptor atom and the hole occurred in a valence band. 
The hole in the otherwise full valence band must be treated as a fixed initial particle 
state with momentum $-\,\bm{k}_e$ according to the principles of scattering theory.

However, as a preparation for this, it is useful to first consider the case of 
a highly excited acceptor or donor state $|i\rangle$ with energy $E_i>E_n(\bm{k}_e)$, 
which overlaps or is even above a conduction band $E_n(\bm{k}_e)$.

\begin{figure}[htb]
\scalebox{0.9}{\includegraphics{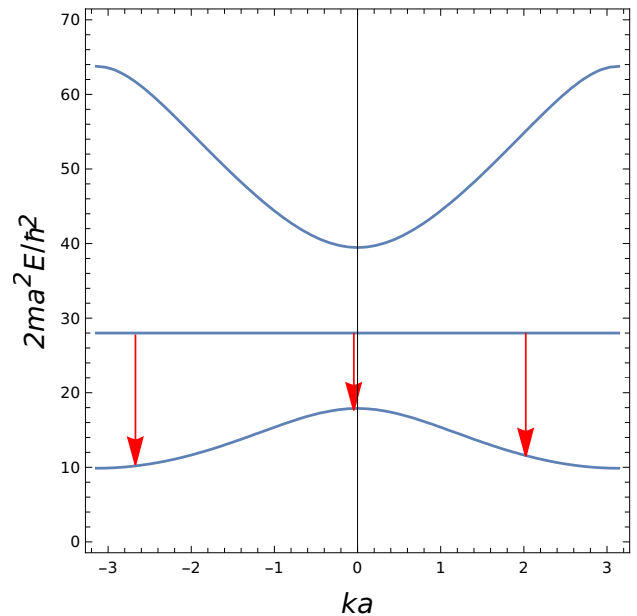}}
\caption{\label{fig:dcemit}
Electronic transitions from a $\bm{k}_e$-independent discrete energy level
into a lower lying energy band. The two bands depicted are the second and third energy
bands in a Kronig-Penney model $V(x)=\mathcal{W}\sum_n\delta(x-na)$
with parameter $m\mathcal{W}a/\hbar^2=-\,7$.}
\end{figure}

Continuous electron states are often labeled either through a
wave vector $\bm{k}_e$, $|\bm{k}_e\rangle$, 
or through continuous energy eigenvalues $E$ and discrete degeneracy 
indices $\nu$, $|E,\nu\rangle$, see Appendix A. Examples of transitions from
a discrete energy level into the $\bm{k}_e$-dependent levels of a lower lying energy band
are depicted in Fig.~\ref{fig:dcemit}. 

Transitions can occur into the whole energy band, i.e.~into Bloch states with
arbitrary wave vector $\bm{k}_e$, because the Bloch wave vectors are confined
to the Brillouin zone $k_e\le\pi/a$ and the Bloch wavelengths $\lambda_e\ge 2a$
imply that the plane wave factors vary slowly over the extent of atomic wave functions.

The appearance of the density of states 
$\tilde{\varrho}(E,\nu)$ in the contributions from the continuous states to the completeness
relations (\ref{eq:completeE1}) informs the appearance of $\tilde{\varrho}(E,\nu)$ in transition 
rates involving continuous states. 

The differential emission rate for photons with polarization $\bm{\epsilon}_\gamma$
and frequency $ck=\omega_{if}$
into a solid angle $d\Omega$ due to transitions from the discrete initial
state $|i\rangle$ into states $|f\rangle=|E_f,\nu_f\rangle$ in
an energy range $[E_f,E_f+dE_f]$ below $E_i$ is
\begin{equation}\label{eq:emdcE}
\frac{d\Gamma_{fi}}{d\Omega}=\frac{\alpha_S}{2\pi c^2}\omega_{if}^3
|\langle f|\bm{\epsilon}^+_\gamma\cdot\op{x}|i\rangle|^2\tilde{\varrho}_f(E_f)dE_f.
\end{equation}
Please note that $\tilde{\varrho}_f(E_f)\equiv\tilde{\varrho}(E_f,\nu_f)$ only provides the 
density of states $|E_f,\nu_f\rangle$ with fixed degeneracy indices $\nu_f$,
see Appendix A.

Eq.~(\ref{eq:emdcE}) must be summed over the degeneracy indices $\nu_f$ if we are 
interested in the differential emission rate of all photons with frequency $ck$
in a range $c[k,k+dk]$, which occur due to the decay of the state $|i\rangle$,
\begin{equation}
\label{eq:emdcE2}
\frac{d\Gamma_{i}}{d\Omega dk}=\frac{\alpha_S}{2\pi}\hbar c^2 k^3
\sum_{\nu_f}|\langle f|\bm{\epsilon}^+_\gamma\cdot\op{x}|i\rangle|^2\tilde{\varrho}_f(E_i-\hbar ck).
\end{equation}

Continuous electron states in materials can be described as
Bloch states with wave vectors $\bm{k}_e$ and energy bands $E_n(\bm{k}_e)$,
\begin{equation}
|f\rangle=|n,\bm{k}_e\rangle=\sqrt{\frac{V}{8\pi^3}}\exp(\mathrm{i}\bm{k}_e\cdot\op{x})
|u_n(\bm{k}_e)\rangle.
\end{equation}
Here $V$ is the volume of the Wigner-Seitz cell and $|u_f\rangle=|u_n(\bm{k}_e)\rangle$ 
is the periodic Bloch factor of the final electron state. 

The differential emission rate for transitions from an initial discrete state $|i\rangle$ into 
states with wave vector $\bm{k}_e$ and energy $E_n(\bm{k}_e)$ in a volume $d^3\bm{k}_e$ in the 
Brillouin zone is
\begin{eqnarray} \label{eq:emdck}
\frac{d\Gamma_{fi}}{d\Omega}&=&\frac{\alpha_S}{2\pi c^2}\omega_{if}^3
|\langle n,\bm{k}_e|\bm{\epsilon}^+_\gamma\cdot\op{x}|i\rangle|^2 d^3\bm{k}_e.
\end{eqnarray}
To integrate this over the Brillouin zone, we use 
\begin{equation}\label{eq:kparallel}
d^3\bm{k}_e\to d^2\bm{k}_{e\|}\frac{dE_e}{|\partial E_n(\bm{k}_e)/\partial\bm{k}_e|},
\end{equation}
where $d^2\bm{k}_{e\|}$ is an integration measure on the constant energy surface
$E_n(\bm{k}_e)=E_e$. Since $E_e=E_i-\hbar ck$, we have $|dE_e|=\hbar cdk$, and we get the
analogue of Eq.~(\ref{eq:emdcE2}) for transition from the discrete state $|i\rangle$
into a lower-lying energy band $E_n(\bm{k}_e)$,
\begin{eqnarray} \nonumber
\frac{d\Gamma_{i}}{d\Omega dk}&=&\frac{\alpha_S}{2\pi}\hbar c^2 k^3
\\ \label{eq:emdcE2k}
&&\times\!\!\int_{E_n(\bm{k}_e)=E_i-\hbar ck}\!\!d^2\bm{k}_{e\|}
\frac{|\langle n,\bm{k}_e|\bm{\epsilon}^+_\gamma\cdot\op{x}|i\rangle|^2}{
|\partial E_n(\bm{k}_e)/\partial\bm{k}_e|}.
\end{eqnarray}

The unpolarized differential emission rates into all directions from Eqs.~(\ref{eq:emdcE2})
and (\ref{eq:emdcE2k}) are
\begin{equation}\label{eq:emdc3}
\frac{d\tilde{\Gamma}_{i}}{dk}=\frac{4\alpha_S}{3}\hbar c^2 k^3
\sum_{\nu_f}|\langle f|\op{x}|i\rangle|^2\tilde{\varrho}_f(E_i-\hbar ck)
\end{equation}
and 
\begin{eqnarray} \nonumber
\frac{d\tilde{\Gamma}_{i}}{dk}&=&\frac{4\alpha_S}{3}\hbar c^2 k^3
\\ \label{eq:emdck2}
&&\times\!\!\int_{E_n(\bm{k}_e)=E_i-\hbar ck}\!d^2\bm{k}_{e\|}
\frac{|\langle n,\bm{k}_e|\op{x}|i\rangle|^2}{
|\partial E_n(\bm{k}_e)/\partial\bm{k}_e|},
\end{eqnarray}
respectively.

Recoil of the emitting atom does not change the form of the
equations (\ref{eq:emdcE}-\ref{eq:emdck2})
but can shift photon frequencies by small amounts, see Appendix C.
Recoil is even more strongly suppressed through the embedding of the
emitting acceptor atom in the crystal lattice, which amounts
to $m_p\to\infty$ in Appendix C.

The case of a conduction band $E_n(\bm{k}_e)$ overlapping or below a highly excited 
acceptor or donor state $|i\rangle$ is not a common situation in solid state spectroscopy. 
Instead, we would rather 
encounter the case of a filled electron acceptor state $|i\rangle$ above a 
valence band $E_n(\bm{k}_e)$ with a hole in the electron state $|f\rangle=|n,\bm{k}_e\rangle$.
In this case we have to consider the hole rather as a continuous initial
state in the problem with a differential current density \cite{Note2} 
\begin{eqnarray}\nonumber
  d\bm{{J}}_h(\bm{k}_e,\bm{x})&=&-\,\psi_n^+(\bm{k}_e,\bm{x})\frac{\hbar}{2\mathrm{i}m}
  \frac{\partial\psi_n(\bm{k}_e,\bm{x})}{\partial\bm{x}}d^3\bm{k}_e
\\ \label{eq:dJh}
&&+\,\frac{\hbar}{2\mathrm{i}m}\frac{\partial\psi_n^+(\bm{k}_e,\bm{x})}{\partial\bm{x}}
\psi_n(\bm{k}_e,\bm{x})d^3\bm{k}_e.
\end{eqnarray}
Please note that the hole is still described by energy eigenvalues $E_n(\bm{k}_e)$
and the corresponding Bloch energy eigenfunctions,
i.e.~we are still dealing with a particle of mass $m_e$ moving through the full periodic
lattice potential without any parabolic band or effective mass approximation. For the
evaluation of the electron recombination amplitude from the initially occupied acceptor state,
the hole is also still described by the electronic annihilation operator $a_{n,s}(\bm{k}_e)$
that generates the vacancy in the valence band: In the hole picture, we go from a depleted
Fermi ground state $a_{n,s}(\bm{k}_e)\bm{|}\Omega\bm{\rangle}$ (tensored with the occupied
atomic acceptor state) to a full Fermi ground
state $\bm{|}\Omega\bm{\rangle}$. However, after evaluation of all the operators, 
this still reduces to the differential transition rate (\ref{eq:emdck}).

The current density (\ref{eq:dJh}) depends on the position $\bm{x}$ in the Wigner-Seitz cell
due to the Bloch factors.
Averaging over the Wigner-Seitz cell yields an analog of the $\bm{x}$-independent current density
of free particles,
\begin{equation}
\frac{d\bm{j}_h(\bm{k}_e)}{d^3\bm{k}_e}=\frac{1}{V}\int_V\!d^3\bm{x}\,
\frac{d\bm{{J}}_h(\bm{k}_e,\bm{x})}{d^3\bm{k}_e}.
\end{equation}
This can be used to divide the differential photon emission
rate $d\Gamma_{fi}/(d\Omega d^3\bm{k}_e)$
from Eq.~(\ref{eq:emdck}) by the norm $|d\bm{j}_h/d^3\bm{k}_e|$
of the differential current density, and
integration over emission directions yields a hole capture cross section,
\begin{equation}\label{eq:holecapture}
\sigma_{fi}(\bm{k}_e)=\frac{4\alpha_S}{3c^2}\omega_{if}^3
\frac{|\langle n,\bm{k}_e|\op{x}|i\rangle|^2}{|d\bm{j}_h/d^3\bm{k}_e|}.
\end{equation}
The photon emission rate due to hole capture from a hole current
density $d\bm{j}_h(\bm{k}_e)/d^3\bm{k}_e$
with energy $E_f(\bm{k}_e)$ of the empty electron states is then 
the radiative hole capture rate,
\begin{equation}
  \tilde{\Gamma}_{fi}=\int_{\mathcal{B}}\!d^3\bm{k}_e\,\sigma_{fi}(\bm{k}_e)
  \left|d\bm{j}_h(\bm{k}_e)/d^3\bm{k}_e\right|,
\end{equation}
where the integration is over the Brillouin zone.

\subsection{Emission rates for $\bm{|i\rangle}$ continuous and $\bm{|f\rangle}$ discrete}

Photon emission from donor recombination with an electron from the conduction band
or from annihilation of a core hole with a conduction electron
are examples for these kinds of transitions. The initial Bloch state in the conduction
band $E_n(\bm{k}_e)$ is $|i\rangle=|n,\bm{k}_e\rangle$.

\begin{figure}[htb]
\scalebox{0.9}{\includegraphics{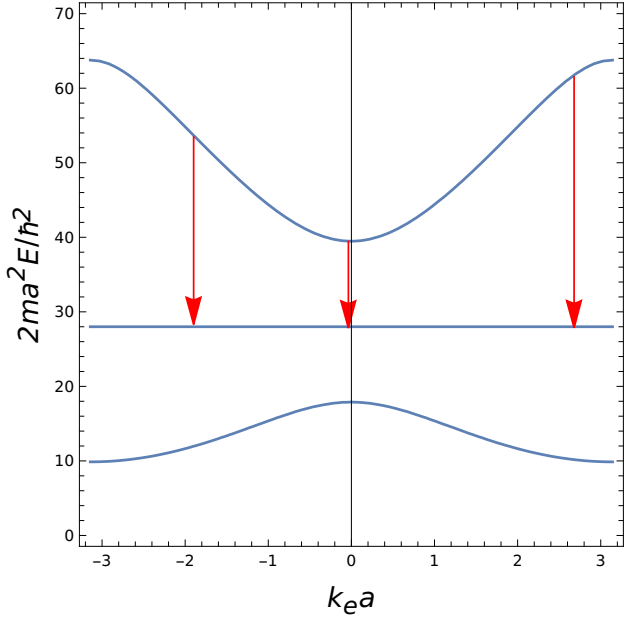}}
\caption{\label{fig:cdemit}
  Electronic transitions from an energy band into the $\bm{k}_e$-independent discrete energy
  level of an atom. The two bands depicted are the second and third energy
bands in a Kronig-Penney model $V(x)=\mathcal{W}\sum_n\delta(x-na)$
with parameter $m\mathcal{W}a/\hbar^2=-\,7$.}
\end{figure}

Examples of transitions from an energy bandy into the $\bm{k}_e$-independent
discrete energy level of an atom (e.g.~from electron-donor recombination)
are depicted in Fig.~\ref{fig:cdemit}. 

Transitions can occur from the whole energy band, i.e.~from Bloch states with
arbitrary wave vector $\bm{k}_e$, because the limit $\lambda_e\ge 2a$ on Bloch wavelengths 
implies that the plane wave factors vary slowly over the extent of atomic wave functions.

The process of radiative electron capture from the conduction band due to a discrete state 
below the conduction band is the mirror process to the radiative hole capture process 
(\ref{eq:holecapture}). This yields an electron capture cross section for electrons with 
differential current density $d\bm{j}_e/d^3\bm{k}_e$,
\begin{equation}\label{eq:electroncapture}
\sigma_{fi}=\frac{4\alpha_S}{3c^2}\omega_{if}^3
\frac{|\langle f|\op{x}|n,\bm{k}_e\rangle|^2}{|d\bm{j}_e/d^3\bm{k}_e|},
\end{equation}
and a corresponding photon emission rate from a differential electron current 
density $d\bm{j}_e/d^3\bm{k}_e$,
\begin{equation}
  \tilde{\Gamma}_{fi}=\int\!d^3\bm{k}_e\,\sigma_{fi}(\bm{k}_e)
  \left|d\bm{j}_e(\bm{k}_e)/d^3\bm{k}_e\right|.
  \end{equation}

The observations from Appendix C about possible shifts in $\omega_{if}$ from
atomic recoil apply here as well, with the note of additional suppression
of recoil effects through embedding of the donor atom in the crystal lattice.

\subsection{Emission rates for both $\bm{|i\rangle}$ and $\bm{|f\rangle}$ continuous}
\label{sec:ecc}

Photon processes involving both continuous initial and continuous final electronic states 
are very different from processes where at least one of the states is discrete.
The presence of at least one discrete state usually requires presence of a participating
atom or molecule, and while the atomic or molecular recoils balance momentum conservation 
in these processes, their impact on the transition rates is limited to a small shift
in photon energies, see Appendix C. This
is why we can effectively treat processes involving discrete electronic states as occurring 
due to interaction with a fixed atom or molecule and the corresponding scattering matrices
then only involve energy conserving $\delta$-functions but no momentum 
conserving $\delta$-functions.

Continuous to continuous transitions, on the other hand, require interband 
transitions and the momentum conserving $\delta$ functions now track changes
in electron momentum instead of atomic recoils. Since 
the wavelength of emitted (or absorbed) photons is much larger than typical lattice
constants, momentum conservation for photon emission and absorption without phonon
assistance leads to direct interband transitions relative to the size
of the Brillouin zone. This is a consequence of the fact that energy bands do not
change much on momentum scales which are very small relative to the size of 
the Brillouin zone. Energy conservation
$\hbar ck=E_i(\bm{k}_{e,i})-E_f(\bm{k}_{e,f})$ therefore simplifies due to
\begin{eqnarray}\nonumber
E_f(\bm{k}_{e,f})&=&E_f(\bm{k}_{e,i}-\bm{k})\simeq E_f(\bm{k}_{e,i})-\bm{k}\cdot
\frac{\partial}{\partial\bm{k}_{e,i}}E_f(\bm{k}_{e,i})
\\ \label{eq:direct}
&\simeq& E_f(\bm{k}_{e,i}),
\end{eqnarray}
i.e.~the energy of the emitted photon effectively corresponds to the energy difference 
between energy bands at the same point in the Brillouin zone.

\begin{figure}[htb]
\scalebox{0.9}{\includegraphics{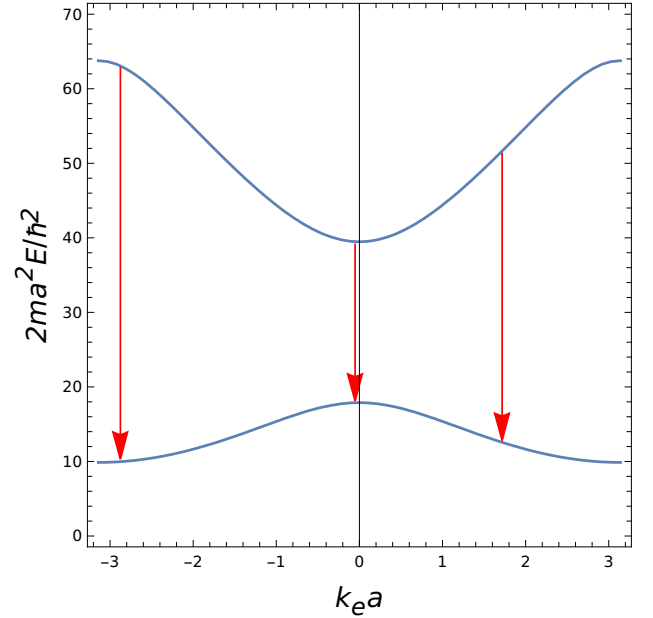}}
\caption{\label{fig:ccemit}
Electronic transitions between the third and second energy
band in a Kronig-Penney model $V(x)=\mathcal{W}\sum_n\delta(x-na)$
with parameter $m\mathcal{W}a/\hbar^2=-\,7$.}
\end{figure}

Here we discuss the case of emission due to transition between
conduction electron states, i.e.~an electron in a higher conduction band
jumps through photon emission into a lower conduction band,
see Fig.~\ref{fig:ccemit} for a schematic.

The initial and final states are Bloch energy eigenstates with energy band indices $n$
and electron wave vectors $\bm{k}_e$,
\begin{eqnarray}\nonumber
\langle\bm{x}|i\rangle&=&\langle\bm{x}|n_i,\bm{k}_{e,i}\rangle=\psi_{n_i}(\bm{k}_{e,i},\bm{x})
\\
&=&\sqrt{\frac{V}{8\pi^3}}\exp(\mathrm{i}\bm{k}_{e,i}\cdot\bm{x})u_{n_i}(\bm{k}_{e,i},\bm{x}),
\end{eqnarray}
 $|f\rangle=|n_f,\bm{k}_{e,f}\rangle$. The orthogonality relation 
\begin{equation}
\langle n,\bm{k}_e|n',\bm{k}'_e\rangle=\delta_{n,n'}\delta(\bm{k}_e-\bm{k}'_e)
\end{equation} 
of the Bloch
states implies orthonormalization in the Wigner-Seitz cell for the Bloch factors 
with the same wave vector,
\begin{eqnarray} \nonumber
\langle u_n(\bm{k}_e)|u_{n'}(\bm{k}_e)\rangle_V&=&\int_V\!d^3\bm{x}\,u^+_n(\bm{k}_e,\bm{x})
u_{n'}(\bm{k}_e,\bm{x})
\\ \label{eq:orthou}
&=&\delta_{n,n'},
\end{eqnarray}
where the integration is over the Wigner-Seitz cell.

The differential photon emission rate in the direction of photon 
momentum $\hat{\bm{k}}$ is 
\begin{eqnarray}\nonumber
\frac{d\Gamma_{fi}}{d\Omega}&=&\frac{\alpha_S}{2\pi c^2}\omega_{if}^3
|\langle f|\bm{\epsilon}^+_\gamma\cdot\op{x}|i\rangle_V|^2\left(\frac{8\pi^3}{V}\right)^2
\\  \label{eq:ecck}
&=&\frac{\alpha_S}{2\pi c^2}\omega_{if}^3
|\langle u_f|\bm{\epsilon}^+_\gamma\cdot\op{x}|u_i\rangle_V|^2.
\end{eqnarray}
The matrix element $\langle f|\bm{\epsilon}^+_\gamma\cdot\op{x}|i\rangle_V$ 
is integrated over the Wigner-Seitz cell, see (\ref{eq:orthou}).

The factor $(8\pi^3/V)^2$ appears in the first line of Eq.~(\ref{eq:ecck})
because on the one hand, the $\bm{k}$-space volume
of a fixed momentum state $|n,\bm{k}_e\rangle$ in a crystal of size
$\mathcal{V}=N_{\bm{R}}V$ is $d^3\bm{k}_e=8\pi^3/N_{\bm{R}}V$, where $N_{\bm{R}}$
is the number of lattice cells.
On the other hand, summation over the locations $\bm{R}$ of lattice cells yields
\begin{equation}\label{eq:sumR}
\sum_{\bm{R}}\exp[\mathrm{i}(\bm{k}_{e,i}-\bm{k}_{e,f}-\bm{k})\cdot\bm{R}]
=\frac{8\pi^3}{V}\delta(\bm{k}_{e,i}-\bm{k}_{e,f}-\bm{k}).
\end{equation}
This factor occurs in the calculation of the scattering matrix element for the photon emission.
The differential photon emission rate from the lattice therefore incurs a factor
\begin{eqnarray}\nonumber
\left[\frac{8\pi^3}{V}\delta(\bm{k}_{e,i}-\bm{k}_{e,f}-\bm{k})\right]^2d^3\bm{k}_{e,i}
&=&\frac{8\pi^3}{V}\delta(\bm{k}_{e,i}-\bm{k}_{e,f}-\bm{k})
\\
&&\times N_{\bm{R}}\frac{8\pi^3}{N_{\bm{R}}V}.
\end{eqnarray}
The momentum conserving $\delta$-function is absorbed by integration over the momentum
of the final electron state.

The corresponding unpolarized emission rate into all directions is
\begin{equation}\label{eq:ecck2}
\tilde{\Gamma}_{fi}=\frac{4\alpha_S}{3c^2}\omega_{if}^3
|\langle u_f|\op{x}|u_i\rangle_V|^2.
\end{equation}

It should be emphasized that Eqs.~(\ref{eq:ecck},\ref{eq:ecck2}) apply to transitions
between conduction bands, but not to a conduction electron filling a valence band hole.
Due to kinematic constraints, photon emission from electron-hole recombination between
conduction and valence band states should be impacted by higher order processes like
phonon assistance, emission of Auger electrons, trapping of particles,
or two-photon emission. It is therefore
outside of the scope of this review, see also Appendix \ref{sec:eh}.

\section{Photon absorption $\bm{|i,\gamma\rangle\to |f\rangle}$}
\label{sec:abs}

In the case of photon absorption, the differential
absorption rate $d\mathcal{A}_{fi}/d^3\bm{k}$ for photons with momentum $\hbar\bm{k}$
is divided by the differential current density $|d\bm{j}_\gamma/d^3\bm{k}|$
of the incoming photons to yield a measure for the efficiency of photon absorption 
through a transition $|i,\gamma\rangle\to |f\rangle$. 
This yields the absorption cross
section $\sigma_{fi}=(d\mathcal{A}_{fi}/d^3\bm{k})/|d\bm{j}_\gamma/d^3\bm{k}|$.

The absorption cross section
\begin{equation}
  \sigma_{fi}(\bm{\epsilon}_{\alpha})
  =[d\mathcal{A}_{fi}(\bm{\epsilon}_{\alpha})/d^3\bm{k}]/|d\bm{j}_\gamma(\bm{\epsilon}_{\alpha})/d^3\bm{k}|
  \end{equation}
 for photons with polarization $\bm{\epsilon}_{\alpha}$ 
depends on $|\langle f|\bm{\epsilon}_{\alpha}\cdot\op{x}|i\rangle|^2$ and therefore on angles
between the vectors $\bm{\epsilon}_{\alpha}$ and $\langle f|\op{x}|i\rangle$.
Averaging over those angles yields
\begin{equation}\label{eq:angleav1}
\frac{1}{4\pi}\int\!d\Omega\,|\langle f|\bm{\epsilon}_{\gamma}\cdot\op{x}|i\rangle|^2
=\frac{1}{3}|\langle f|\op{x}|i\rangle|^2,
\end{equation}
thus removing the dependence on polarization, see also 
Eqs.~(\ref{eq:total1}) and (\ref{eq:avproof1}-\ref{eq:avproof4}).

On the other hand, normalization by the unpolarized photon current
density $d\bm{j}_\gamma/d^3\bm{k}=\sum_\alpha d\bm{j}_\gamma(\bm{\epsilon}_{\alpha})/d^3\bm{k}$
and $d\bm{j}_\gamma(\bm{\epsilon}_{1})/d^3\bm{k}=d\bm{j}_\gamma(\bm{\epsilon}_{2})/d^3\bm{k}$ for
unpolarized currents
implies that the unpolarized cross section is the average of the polarized cross
sections,
\begin{equation}
  \sigma_{fi}=\frac{1}{|d\bm{j}_\gamma/d^3\bm{k}|}
  \sum_{\alpha=1}^2\frac{d\mathcal{A}_{fi}(\bm{\epsilon}_{\alpha})}{d^3\bm{k}}
  =\frac{1}{2}\sum_{\alpha=1}^2\sigma_{fi}(\bm{\epsilon}_{\alpha}).
\end{equation}

Since angle averaging removes polarization dependence (\ref{eq:angleav1}),
the angle averaged absorption cross section
\begin{equation}
\overline{\sigma}_{fi}=\frac{1}{4\pi}\int\!d\Omega\,\sigma_{fi}
\end{equation}
has the same value for polarized and unpolarized photons.
 Furthermore, Eq.~(\ref{eq:angleav1})
implies that $\overline{\sigma}_{fi}$ is gotten from $\sigma_{fi}$ through the substitution
\begin{equation}\label{eq:angleav2}
|\langle f|\bm{\epsilon}_{\gamma}\cdot\op{x}|i\rangle|^2\to\frac{1}{3}|\langle f|\op{x}|i\rangle|^2.
\end{equation}
An example is provided in Eqs.~(\ref{eq:absdd1},\ref{eq:absdd2}). Due to the simple substitution rule
(\ref{eq:angleav2}) we will only write down the polarized absorption cross sections $\sigma_{fi}$
in other cases.

If we are interested in the aborption cross sections for all photons with frequency $\omega_{fi}$
due to absorption from the initial state $|i\rangle$, we need to sum the equations for
the absorption cross sections $\sigma_{fi}$ over the degeneracy indices $\nu_f$ of the final 
states,
\begin{equation}
\sigma_i\Big|_{ck=\omega_{fi}}=\sum_{\nu_f}\sigma_{fi}.
\end{equation}

\subsection{Absorption cross sections for both $\bm{|i\rangle}$ and $\bm{|f\rangle}$ discrete}

The absorption cross section for photons with polarization $\bm{\epsilon}_\gamma$
and momentum $\hbar\bm{k}$ is
\begin{equation}\label{eq:absdd1}
\sigma_{fi}=4\pi^2\alpha_S\omega_{fi}|\langle f|\bm{\epsilon}_\gamma\cdot\op{x}|i\rangle|^2
\delta(\omega_{fi}-ck).
\end{equation}
The angle averaged absorption cross section for polarized or unpolarized
photons follows immediately from (\ref{eq:angleav1}),
\begin{equation}\label{eq:absdd2}
\overline{\sigma}_{fi}=
\frac{4\pi^2}{3}\alpha_S\omega_{fi}|\langle f|\op{x}|i\rangle|^2\delta(\omega_{fi}-ck).
\end{equation}
Eqs.~(\ref{eq:absdd1},\ref{eq:absdd2}) follow in the displayed forms
directly from the scattering matrix in first order time-dependent perturbation theory.
However, closer examination of the transition rates through inclusion of the reduction
of the initial state in a coupled set of rate equations \cite{LorentzWW,Merzbacher}
or through the resolvent operator \cite{Messiah,Scala} replaces the energy
conserving $\delta$ function in Eqs.~(\ref{eq:absdd1},\ref{eq:absdd2})
with a Lorentz profile $\Delta_\Gamma(\omega_{fi}-ck)$ of width $2\Gamma$, 
\begin{equation} \label{eq:lorentz}
\delta(\omega_{fi}-ck)\to\Delta_\Gamma(\omega_{fi}-ck)
=\frac{1}{\pi}\frac{\Gamma}{(\omega_{fi}-ck)^2+\Gamma^2}.
\end{equation}
We have to be careful, however, to note that the direct substitution (\ref{eq:lorentz})
only works as an approximation for narrow spectral lines, $\Gamma\ll\omega_{fi}$, when we
are always at or near resonance. Otherwise an additional factor $\omega_{fi}/ck$ would appear
in Eqs.~(\ref{eq:absdd1},\ref{eq:absdd2}) because there would be the factor $\omega_{fi}^2$
arising from Eq.~(\ref{eq:l2v}),
$\langle f|\op{p}|i\rangle=\mathrm{i}m\omega_{fi}\langle f|\op{x}|i\rangle$,
and there would be a factor $1/ck$ arising from a factor $1/\sqrt{ck}$
in the mode expansion of the vector potential. Furthermore, wide lines may
be caused by other line broadening effects besides lifetime broadening and
a Lorentzian profile may not be appropriate anymore.

\subsection{Absorption cross sections for $\bm{|i\rangle}$ discrete and $\bm{|f\rangle}$ continuous}
\label{subsec:ad2c}

\begin{figure}[htb]
\scalebox{0.9}{\includegraphics{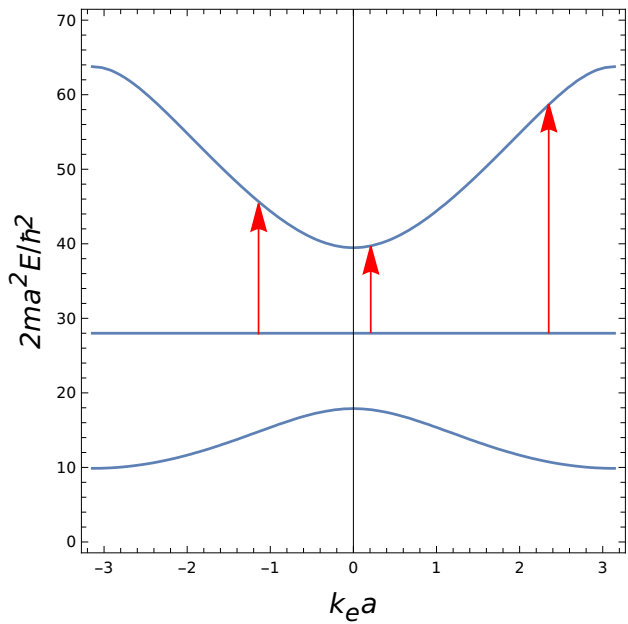}}
\caption{\label{fig:dcabsorb}
Electronic transitions from a $\bm{k}_e$-independent discrete donor level
into a higher energy band. The two bands depicted are the second and third energy
bands in a Kronig-Penney $V(x)=\mathcal{W}\sum_n\delta(x-na)$
with parameter $m\mathcal{W}a/\hbar^2=-\,7$.}
\end{figure}

This case applies e.g.~to ionization of an atom or of a donor in a semiconductor,
see Fig.~\ref{fig:dcabsorb} for a schematic.

Transitions can occur into the whole energy band, i.e.~into Bloch states with
arbitrary wave vector $\bm{k}_e$, because the limit $\lambda_e\ge 2a$ on Bloch wavelengths 
implies that the plane wave factors vary slowly over the extent of atomic wave functions.

The polarized absorption cross section for transition into states $|f\rangle=|E_f,\nu_f\rangle$ 
in the energy range $[E_f,E_f+dE_f]$ and with discrete degeneracy indices $\nu_f$ is
\begin{equation}\label{eq:absdc1}
\sigma_{fi}=4\pi^2\alpha_S\hbar ck
\left[\tilde{\varrho}_f(E_f)|\langle f|\bm{\epsilon}_\gamma\cdot\op{x}|i\rangle|^2\right]_{E_f=E_i+\hbar ck}.
\end{equation}

On the other hand, if $|i\rangle$ is a discrete donor state and
$|f\rangle=|n,\bm{k}_e\rangle$ are states in the conduction band $E_n(\bm{k}_e)$,
the differential absorption cross section for polarized photons of frequency $ck$ from 
occupied donor states $|i\rangle$ due to electron excitation into the conduction band is
\begin{equation}\label{eq:absB1}
d\sigma_{fi}=4\pi^2\alpha_S\omega_{fi}|\langle n,\bm{k}_e|\bm{\epsilon}_\gamma\cdot\op{x}|i\rangle|^2
\delta(\omega_{fi}-ck)d^3\bm{k}_e.
\end{equation}
The total contribution from the conduction band $E_n(\bm{k}_e)$
to the absorption cross section for polarized photons of frequency $ck$ is
\begin{eqnarray}\nonumber
\sigma_{ni}&=&4\pi^2\alpha_S\hbar ck
\\ \nonumber
&&\times\!\int_{\mathcal{B}}\!d^3\bm{k}_e\,
|\langle n,\bm{k}_e|\bm{\epsilon}_\gamma\cdot\op{x}|i\rangle|^2
\delta(E_n(\bm{k}_e)-E_i-\hbar ck)
\\ \nonumber
&=&4\pi^2\alpha_S\hbar ck
\\ \label{eq:absB3}
&&\times\!\int_{E_n(\bm{k}_e)=E_i+\hbar ck}\!d^2\bm{k}_{e\|}\,
\frac{|\langle n,\bm{k}_e|\bm{\epsilon}_\gamma\cdot\op{x}|i\rangle|^2}{
|\partial E_n(\bm{k}_e)/\partial\bm{k}_e|}.
\end{eqnarray}
Here $d^2\bm{k}_{e\|}$ is an integration measure along the constant energy
surface $E_n(\bm{k}_e)=E_i+\hbar ck$ in the Brillouin zone $\mathcal{B}$. 

If the matrix element $|\langle n,\bm{k}_e|\bm{\epsilon}_\gamma\cdot\op{x}|i\rangle|^2$
is approximately constant over the constant energy surface, we find an equation very
similar to (\ref{eq:absdc1}),
\begin{eqnarray}\nonumber
\sigma_{ni}&=& 4\pi^2\alpha_S\hbar ck
\\ \label{eq:absB4}
&&\times\!
\left[\varrho_{n,V}(E_f)|\langle u_n(\bm{k}_e)|
\bm{\epsilon}_\gamma\cdot\op{x}|i\rangle|^2\right]_{E_f=E_i+\hbar ck},
\end{eqnarray}
with the density of states $\varrho_{n,V}(E)$ in the energy band $E_n(\bm{k}_e)$,
\begin{equation}
\label{eq:rhonE}
\varrho_{n,V}(E)
=\frac{V}{8\pi^3}\int_{E_n(\bm{k}_e)=E}
\frac{d^2\bm{k}_{e\|}}{|\partial E_n(\bm{k}_e)/\partial\bm{k}_e|}.
\end{equation}
Here we used that the discrete initial state $|i\rangle$ should be localized within a Wigner-Seitz 
cell and therefore $\exp(\mathrm{i}\bm{k}_e\cdot\bm{x})\simeq 1$ in the matrix element.

Intuitively, the assumption of approximately 
constant factor $|\langle u_n(\bm{k}_e)|\bm{\epsilon}_\gamma\cdot\op{x}|i\rangle|^2$ across the 
constant energy surface $E_n(\bm{k}_e)=E_i+\hbar ck$ could be justified if that energy surface
is small compared to the typical area $4\pi^2/V^{2/3}$ of the Brillouin zone.
The integral in (\ref{eq:emdcE2k}) then spans a relatively small surface area in the Brillouin zone 
such that the Bloch factor does not vary a lot with $\bm{k}_e$.

 For the interpretation of Eq.~(\ref{eq:rhonE}), we note that
the local number of spin-polarized electron states in the phase space volume 
$d^3\bm{x}d^3\bm{k}_e$ and in the energy band $E_n(\bm{k}_e)$ is 
\begin{eqnarray}\nonumber
dN_n(\bm{k}_e,\bm{x})&=&|\langle\bm{x}|n,\bm{k}_e\rangle|^2 d^3\bm{x}d^3\bm{k}_e
\\ \label{eq:dNn}
&=&\frac{V}{8\pi^3}|\langle\bm{x}|u_n(\bm{k}_e)\rangle|^2
d^3\bm{x}d^3\bm{k}_e.
\end{eqnarray}
The density $\varrho_{n,V}(E)$ is therefore the contribution from the energy band $E_n(\bm{k}_e)$
to the density of spin-polarized electron states in the energy scale and in the Wigner-Seitz cell,
i.e.
\begin{equation}
N^{(n)}_{[E_1,E_2]}=\int_{E_1}^{E_2}\!dE\,\varrho_{n,V}(E)
\end{equation}
is the contribution from the energy band $E_n(\bm{k}_e)$ to the number of spin-polarized 
electron states in the Wigner-Seitz cell and with energies $E_1\le E\le E_2$.

Eq.~(\ref{eq:absdc1}) is exact (within the limits of first-order perturbation theory
and dipole approximation) since we assumed fixed discrete degeneracy 
indices $\nu_f$ in the final continuous state, whereas here we have continuous degeneracy 
indices $\bm{k}_{e\|}$ tangential to the surface of constant energy, and we integrated
over those degeneracy indices while ignoring 
the factor $|\langle n,\bm{k}_e|\bm{\epsilon}_\gamma\cdot\op{x}|i\rangle|^2$. 
The density of states $\tilde{\varrho}_f(E_f)$ in the exact 
equation (\ref{eq:absdc1}) is therefore a density of states for fixed discrete degeneracy 
indices $\nu_f$, whereas the density of states $\varrho_{n,V}(E_f)$ in the approximate
equation (\ref{eq:absB4}) is integrated over degeneracies.
On the other hand, summation over the discrete degeneracy indices in (\ref{eq:absdc1}) would 
yield the same result as the integration in (\ref{eq:absB3}).

\subsection{Absorption cross sections for $\bm{|i\rangle}$ continuous and $\bm{|f\rangle}$ discrete}

\begin{figure}[htb]
\scalebox{0.9}{\includegraphics{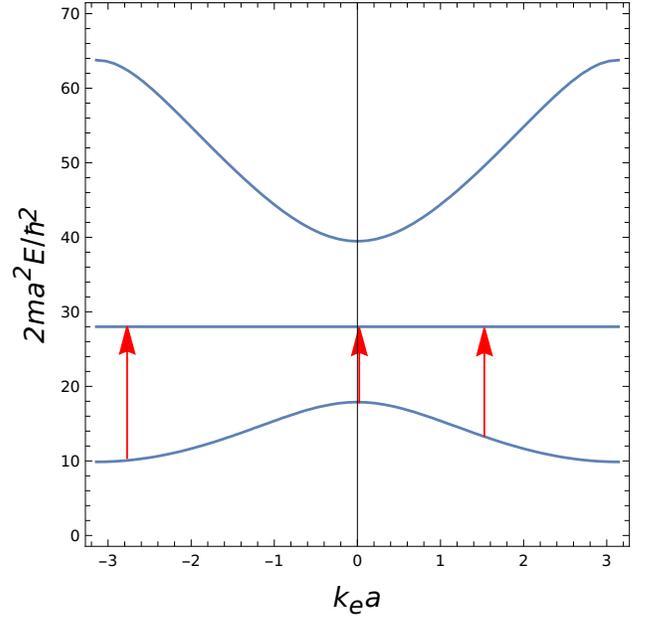}}
\caption{\label{fig:cdabsorb}
Electronic transitions from an energy band into a $\bm{k}_e$-independent discrete
acceptor level. The two bands depicted are the second and third energy
bands in a Kronig-Penney $V(x)=\mathcal{W}\sum_n\delta(x-na)$
with parameter $m\mathcal{W}a/\hbar^2=-\,7$.}
\end{figure}

This applies e.g.~to ionization of an acceptor due to acceptance of a valence band electron,
see Fig.~\ref{fig:cdabsorb} for a schematic.

Transitions can occur from the whole energy band, i.e.~from initial Bloch states with
arbitrary wave vector $\bm{k}_e$, because the limit $\lambda_e\ge 2a$ on Bloch wavelengths
implies that the plane wave factors vary slowly over the extent of atomic wave functions.

The polarized absorption cross section for transition from initial 
states $|i\rangle=|E_i,\nu_i\rangle$ in the energy range $[E_i,E_i+dE_i]$ is
\begin{equation}\label{eq:sficd1}
\sigma_{fi}=4\pi^2\alpha_S\hbar ck
\left[\tilde{\varrho}_i(E_i)|\langle f|\bm{\epsilon}_\gamma\cdot\op{x}|i\rangle|^2\right]_{E_f=E_i+\hbar ck}.
\end{equation}

Eq.~(\ref{eq:sficd1}) assumes discrete degeneracy indices $\nu_i$ and
$\tilde{\varrho}_i(E_i)\equiv\tilde{\varrho}(E_i,\nu_i)$.
On the other hand, if the initial state is a Bloch state $|i\rangle=|n,\bm{k}_e\rangle$
in an energy band $E_n(\bm{k}_e)$, the differential absorption cross section from
transitions into the discrete acceptor state $|f\rangle$ is
\begin{equation}\label{eq:absB2}
d\sigma_{fi}=4\pi^2\alpha_S ck|\langle f|\bm{\epsilon}_\gamma\cdot\op{x}|n,\bm{k}_e\rangle|^2
\delta(\omega_{fi}-ck)d^3\bm{k}_e.
\end{equation}
The contribution from the whole energy band to the absorption cross section is therefore
\begin{eqnarray}\nonumber
\sigma_{fn}&=&4\pi^2\alpha_S\hbar ck
\\ \label{eq:absB2b}
&&\times\!\int_{E_n(\bm{k}_e)=E_i}\!d^2\bm{k}_{e\|}
\frac{|\langle f|\bm{\epsilon}_\gamma\cdot\op{x}|n,\bm{k}_e\rangle|^2}{
|\partial E_n(\bm{k}_e)/\partial\bm{k}_e|}.
\end{eqnarray}
Just as in the previous subsection, the assumption of approximately constant matrix element across
the constant energy surface yields an approximation that resembles Eq.~(\ref{eq:sficd1}),
\begin{eqnarray}\nonumber
\sigma_{fn}&=& 4\pi^2\alpha_S\hbar ck
\\ \label{eq:absB2c}
&&\times\!
\left[\varrho_{n,V}(E_i)|\langle f|\bm{\epsilon}_\gamma\cdot\op{x}|u_n(\bm{k}_e)\rangle|^2
\right]_{E_f=E_i+\hbar ck},
\end{eqnarray}
with the density of states (\ref{eq:rhonE}).

Eq.~(\ref{eq:absB2b}) includes a sum over the continuous degeneracy indices corresponding to the 
integration measure $d^2\bm{k}_{e\|}$ on the constant energy surface $E_n(\bm{k}_e)=E_i$.
This corresponds to summation over the discrete degeneracy indices $\nu_i$
in Eq.~(\ref{eq:sficd1}).

\subsection{Absorption cross sections for both $\bm{|i\rangle}$ and $\bm{|f\rangle}$ continuous}

\begin{figure}[htb]
\scalebox{0.9}{\includegraphics{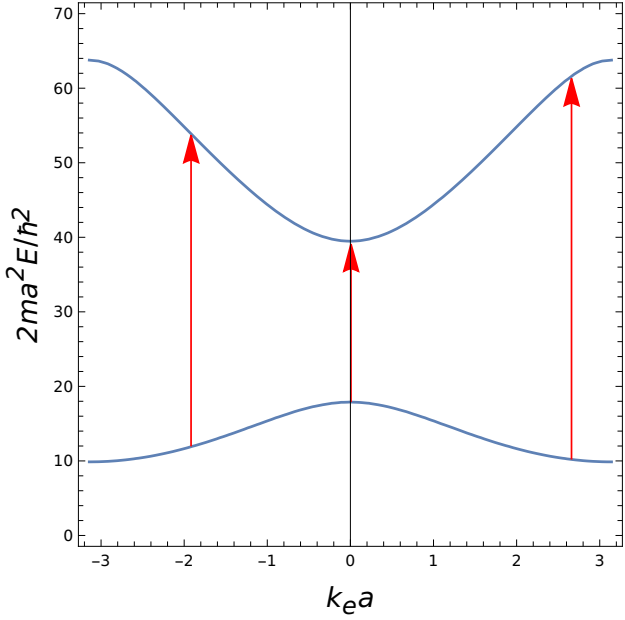}}
\caption{\label{fig:ccabsorb}
Electronic transitions between the second and third energy
band in a Kronig-Penney model $V(x)=\mathcal{W}\sum_n\delta(x-na)$
with parameter $m\mathcal{W}a/\hbar^2=-\,7$.}
\end{figure}

This applies to absorption due to interband transitions, see Fig.~\ref{fig:ccabsorb} for
a schematic. The same remarks
as in subsection \ref{sec:ecc} concerning momentum conservation apply.

The contribution from the transition $|i\rangle=|n,\bm{k}_e\rangle\to |f\rangle=|n',\bm{k}_e\rangle$ 
to the absorption cross section for polarized photons of frequency $ck$ is 
\begin{equation}\label{eq:sigmaBB1}
\sigma_{fi}=4\pi^2\alpha_S ck\delta(\omega_{fi}-ck)
|\langle u_f|\bm{\epsilon}_\gamma\cdot\op{x}|u_i\rangle_V|^2,
\end{equation}
where $V$ is the volume of the Wigner-Seitz cell and $|u_i\rangle$ or $|u_f\rangle$ are the 
Bloch factors of the initial and final states, respectively, see Eq.~(\ref{eq:blochpsi}). 
The matrix element $\langle u_f|\bm{\epsilon}_\gamma\cdot\op{x}|u_i\rangle_V$ is integrated over 
the Wigner-Seitz cell.

The absorption cross section per lattice cell from the interband 
transition $E_n(\bm{k}_e)\to E_{n'}(\bm{k}_e)$ is
\begin{eqnarray}\nonumber
\sigma_{n'n}&=&\frac{V}{8\pi^3}\int\!d^3\bm{k}_e\,\sigma_{fi}
\\ \nonumber
&=&\frac{\alpha_S}{2\pi}\hbar ckV\int_{E_{n'}(\bm{k}_e)-E_n(\bm{k}_e)=\hbar ck}\!d^2\bm{k}_{e\|}
\\
&&\times\frac{
|\langle u_{n'}(\bm{k}_e)|\bm{\epsilon}_\gamma\cdot\op{x}|u_n(\bm{k}_e)\rangle_V|^2}{
|\partial[E_{n'}(\bm{k}_e)-E_n(\bm{k}_e)]/\partial\bm{k}_e|}.
\end{eqnarray}

Pulling the factor $|\langle u_{n'}(\bm{k}_e)|\bm{\epsilon}_\gamma\cdot\op{x}|u_n(\bm{k}_e)\rangle_V|^2$ 
out of the integral under the assumption that it is approximately constant over the surface
$E_{n'}(\bm{k}_e)-E_n(\bm{k}_e)=\hbar ck$ yields an equation which resembles 
Eqs.~(\ref{eq:absdc1},\ref{eq:sficd1}),
\begin{eqnarray}\nonumber
\sigma_{n'n}&=& 4\pi^2\alpha_S\hbar ck
|\langle u_{n'}(\bm{k}_e)|\bm{\epsilon}_\gamma\cdot\op{x}|u_n(\bm{k}_e)\rangle_V|^2
\\
&&\times
\varrho_{n'n,V}(\hbar ck)
\end{eqnarray}
with the joint density of states
\begin{eqnarray}\nonumber
\varrho_{n'n}(E)&=&\frac{V}{8\pi^3}\int\!d^3\bm{k}_e\,
\delta(E_{n'}(\bm{k}_e)-E_n(\bm{k}_e)-E)
\\ \nonumber
&=&\frac{V}{8\pi^3}\int_{E_{n'}(\bm{k}_e)-E_n(\bm{k}_e)=E}\!d^2\bm{k}_{e\|}
\\ \label{eq:jointrho}
&&\times\frac{1}{
|\partial[E_{n'}(\bm{k}_e)-E_n(\bm{k}_e)]/\partial\bm{k}_e|}.
\end{eqnarray}
The reasoning that led to the interpretation of $\varrho_n(E)$ (\ref{eq:rhonE}) implies that the 
joint density of states (\ref{eq:jointrho}) yields the spin-polarized number $\varrho_{n'n,V}(E)dE$ 
of pairs of electron states in the Wigner-Seitz cell
which are contributed by the energy bands $E_{n'}(\bm{k}_e)$ and $E_{n}(\bm{k}_e)$ 
and satisfy $E_{n'}(\bm{k}_e)-E_n(\bm{k}_e)\in [E,E+dE]$.

\section{Photon scattering $\bm{|i,\gamma\rangle\to |f,\gamma'\rangle}$}
\label{sec:sc}

The differential scattering cross sections for photon 
scattering $|i;\bm{\epsilon}_\gamma,\bm{k}\rangle\to|f;\bm{\epsilon}'_\gamma,\bm{k}'\rangle$
involve sums over intermediate electron
states $|v\rangle$ in the form $\sum_v|v\rangle f(E_v)\langle v|$,
where $f(E_v)=\hbar/(E_v-E_i-\hbar ck)$ or $f(E_v)=\hbar/(E_v-E_i+\hbar ck')$, respectively,
in the two scattering terms that appear in the Kramers-Heisenberg formula, see
Eq.~(\ref{eq:KHdd}) below. The sum over the 
intermediate states is just a shorthand notation for sums over discrete intermediate electron
states $|d\rangle$ (e.g.~donor or acceptor states) and continuous intermediate electron states, 
e.g.~due to transition through intermediate states $|n,\bm{k}_e\rangle$ in energy 
bands $E_n(\bm{k}_e)$,
\begin{eqnarray}\nonumber
\sum_v|v\rangle f(E_v)\langle v|&\equiv&\sum_d|d\rangle f(E_d)\langle d|
\\ \nonumber
&&+\sum_n\!\int\!d^3\bm{k}_e\,|n,\bm{k}_e\rangle f(E_n(\bm{k}_e))\langle n,\bm{k}_e|.
\end{eqnarray}

\subsection{Differential scattering cross section for both $\bm{|i\rangle}$ and $\bm{|f\rangle}$
  discrete}

\begin{figure}[htb]
\scalebox{0.9}{\includegraphics{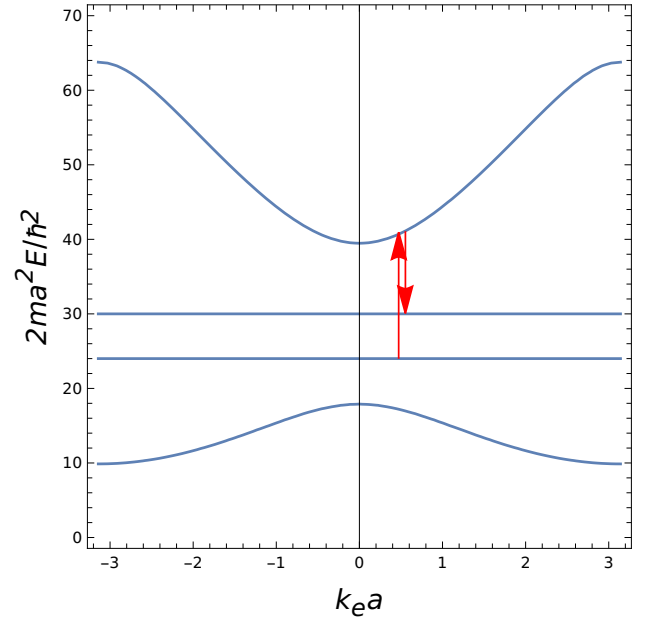}}
\caption{\label{fig:ddscatter}
  Scattering between $\bm{k}_e$-independent discrete atomic energy levels.
  In this case the scattering is resonantly enhanced through
  an intermediate band state.
The depicted bands correspond to the second and third energy
band in a Kronig-Penney model $V(x)=\mathcal{W}\sum_n\delta(x-na)$
with parameter $m\mathcal{W}a/\hbar^2=-\,7$.}
\end{figure}

This case applies to scattering between initial and final bound atomic states, e.g.~core states
in lattice atoms or bound states of acceptor or donor atoms. A schematic involving 
resonantly enhanced
scattering through intermediate energy band states is depicted in Fig.~\ref{fig:ddscatter}

Transitions can involve intermediate states in the whole energy band, i.e.~intermediate
Bloch states with
arbitrary wave vector $\bm{k}_e$, because the limit $\lambda_e\ge 2a$ on Bloch wavelengths 
implies that the plane wave factors vary slowly over the extent of the atomic wave functions.

Minimal coupling $\bm{p}\to\bm{p}+e\bm{A}(\bm{x},t)$
of photons into the Schr\"odinger equation yields the differential photon scattering
cross section \cite{Waller,Dirac,Heitler}
\begin{eqnarray}\nonumber
\frac{d\sigma}{d\Omega}&=&\left(\frac{\alpha_S}{m^2 c}\right)^2\frac{k'}{k}
\\ \nonumber
&&\times\!
\left|m\hbar\delta_{fi}\bm{\epsilon}'_\gamma\cdot\bm{\epsilon}_\gamma
-\sum_v
\frac{\langle f|\bm{\epsilon}'_\gamma\cdot\op{p}|v\rangle\langle v|\bm{\epsilon}_\gamma\cdot\op{p}|i\rangle}{
\omega_{vi}-ck-\mathrm{i}\eta}
\right.
\\ \label{eq:master1p}
&&-\sum_v\left.
\frac{\langle f|\bm{\epsilon}_\gamma\cdot\op{p}|v\rangle\langle v|\bm{\epsilon}'_\gamma
\cdot\op{p}|i\rangle}{\omega_{vi}+ck'-\mathrm{i}\eta}\right|_{\omega_{fi}=ck-ck'}^2\!\!.
\end{eqnarray}
Here we use $\bm{\epsilon}'_\gamma\equiv\bm{\epsilon}^+(\bm{k}')$ for the polarization vector
of the emitted photon while $\bm{\epsilon}_\gamma\equiv\bm{\epsilon}(\bm{k})$ is the polarization vector
of the incident photon. 

The differential scattering cross section with the matrix elements in length form is
\begin{eqnarray}\nonumber
\frac{d\sigma}{d\Omega}&=&\left(\frac{\alpha_S}{c}\right)^2\frac{k'}{k}
\\ \nonumber
&&\times\!
\left|\frac{\hbar}{m}\delta_{fi}\bm{\epsilon}'_\gamma\cdot\bm{\epsilon}_\gamma
+\sum_v\omega_{fv}\omega_{vi}
\frac{\langle f|\bm{\epsilon}'_\gamma\cdot\op{x}|v\rangle\langle v|\bm{\epsilon}_\gamma\cdot\op{x}|i\rangle}{
\omega_{vi}-ck-\mathrm{i}\eta}
\right.
\\ \label{eq:master1}
&&+\sum_v\left.\omega_{fv}\omega_{vi}
\frac{\langle f|\bm{\epsilon}_\gamma\cdot\op{x}|v\rangle\langle v|\bm{\epsilon}'_\gamma\cdot\op{x}|i\rangle}{
\omega_{vi}+ck'-\mathrm{i}\eta}\right|_{\omega_{fi}=ck-ck'}^2\!\!.
\end{eqnarray}
Textbook derivations of the scattering cross sections from the second order expansion of time-evolution 
operators can be found in \cite{Sakurai,Dick}. The structure of the denominators in the
second and third terms in Eqs.~(\ref{eq:master1p},\ref{eq:master1}) follows from the time integrals
in the second order scattering matrix element,
\begin{eqnarray} \nonumber
&&
\int_{-\infty}^\infty\!dt\int_{-\infty}^t\!dt'\,
\exp\!\left[\mathrm{i}(\omega_{fv}+ck')t\right]
\exp\!\left[\mathrm{i}(\omega_{vi}-ck)t'+\eta t'\right]
\\
&&\qquad=-\,2\pi\mathrm{i}\frac{\delta(\omega_{fi}+ck'-ck)}{
\omega_{vi}-ck-\mathrm{i}\eta},
\end{eqnarray}
\begin{eqnarray} \nonumber
&&
\int_{-\infty}^\infty\!dt\int_{-\infty}^t\!dt'\,
\exp\!\left[\mathrm{i}(\omega_{fv}-ck)t\right]
\exp\!\left[\mathrm{i}(\omega_{vi}+ck')t'+\eta t'\right]
\\
&&\qquad=-\,2\pi\mathrm{i}\frac{\delta(\omega_{fi}+ck'-ck)}{
\omega_{vi}+ck'-\mathrm{i}\eta}.
\end{eqnarray}
The shift $\omega_{vi}\to\omega_{vi}-\mathrm{i}\eta$ can be understood as a consequence 
of $E_v\to E_v-\mathrm{i}\hbar\eta$, 
i.e.~$\eta^{-1}$ is the decay time of the intermediate state $|v\rangle$.

The differential scattering cross section $d\sigma/d\Omega$ arises from differential transition rates
into volume elements $d^3\bm{k}'=k'^2 dk'd\Omega$ of final photon states (normalized by incident photon flux), 
after integration over $dk'$ against the energy conserving $\delta$-function $\delta(\omega_{fi}+ck'-ck)$, 
see e.g.~Eqs.~(18.191-18.193) in \cite{Dick}. However, it is also useful to consider the differential
scattering cross section in Wigner-Weisskopf form, $d\sigma/d\Omega dk'$, with substitution of a
Lorentz profile $\Delta_\Gamma(\omega_{fi}+ck'-ck)$ (\ref{eq:lorentz}) for the $\delta$-function 
if lineshapes are resolved \cite{Tulkki,Gelmukhanov1},
\begin{eqnarray}\nonumber
\frac{d\sigma}{d\Omega dk'}&=&\alpha_S^2\frac{k'}{ck}
\Delta_\Gamma(\omega_{fi}+ck'-ck)
\left|\frac{\hbar}{m}\delta_{fi}\bm{\epsilon}'_\gamma\cdot\bm{\epsilon}_\gamma
\right.
\\ \nonumber
&&+\sum_v\omega_{fv}\omega_{vi}
\frac{\langle f|\bm{\epsilon}'_\gamma\cdot\op{x}|v\rangle\langle v|\bm{\epsilon}_\gamma\cdot\op{x}|i\rangle}{
\omega_{vi}-ck-\mathrm{i}\eta}
\\ \label{eq:masterWW}
&&+\sum_v\left.\omega_{fv}\omega_{vi}
\frac{\langle f|\bm{\epsilon}_\gamma\cdot\op{x}|v\rangle\langle v|\bm{\epsilon}'_\gamma\cdot\op{x}|i\rangle}{
\omega_{vi}+ck'-\mathrm{i}\eta}\right|^2\!\!.
\end{eqnarray}

If the scattering is dominated by nearly resonant intermediate states, i.e.~if we have states $|v\rangle$
such that
\begin{equation}\label{eq:res1}
\langle f|\bm{\epsilon}'_\gamma\cdot\op{x}|v\rangle\langle v|\bm{\epsilon}_\gamma\cdot\op{x}|i\rangle\neq 0
\,\wedge\, \omega_{vi}-ck=\omega_{vf}-ck'\simeq 0
\end{equation}
or
\begin{equation}\label{eq:res2}
\langle f|\bm{\epsilon}_\gamma\cdot\op{x}|v\rangle\langle v|\bm{\epsilon}'_\gamma\cdot\op{x}|i\rangle
\neq 0
\,\wedge\,\omega_{vi}+ck'=\omega_{vf}+ck\simeq 0,
\end{equation}
then $\omega_{fv}\omega_{vi}\simeq -\,c^2kk'$
and we can use the Kramers-Heisenberg approximations
\begin{eqnarray}\nonumber
\frac{d\sigma}{d\Omega}&=&\alpha_S^2 c^2 kk'^3
\left|\sum_v
\frac{\langle f|\bm{\epsilon}'_\gamma\cdot\op{x}|v\rangle\langle v|\bm{\epsilon}_\gamma\cdot\op{x}|i\rangle}{
\omega_{vi}-ck-\mathrm{i}\eta}
\right.
\\ \label{eq:KHdd}
&&+\sum_v\left.
\frac{\langle f|\bm{\epsilon}_\gamma\cdot\op{x}|v\rangle\langle v|\bm{\epsilon}'_\gamma\cdot\op{x}|i\rangle}{
\omega_{vi}+ck'-\mathrm{i}\eta}\right|_{\omega_{fi}=ck-ck'}^2
\end{eqnarray}
or
\begin{eqnarray}\nonumber
\frac{d\sigma}{d\Omega dk'}&=&\alpha_S^2 c^3 kk'^3\Delta_\Gamma(\omega_{fi}+ck'-ck)
\\ \nonumber
&&\times\!
\left|\sum_v
\frac{\langle f|\bm{\epsilon}'_\gamma\cdot\op{x}|v\rangle\langle v|\bm{\epsilon}_\gamma\cdot\op{x}|i\rangle}{
\omega_{vi}-ck-\mathrm{i}\eta}
\right.
\\ \label{eq:KHddWW}
&&+\!\left.\sum_v
\frac{\langle f|\bm{\epsilon}_\gamma\cdot\op{x}|v\rangle\langle v|\bm{\epsilon}'_\gamma\cdot\op{x}|i\rangle}{
\omega_{vi}+ck'-\mathrm{i}\eta}\right|^2,
\end{eqnarray}
respectively.

The Kramers-Heisenberg approximation follows directly from the Schr\"odinger equation if, instead 
of minimal coupling, we use a dipole coupling $H_{e\gamma}\simeq e\op{x}\cdot\bm{E}(t)$ for
the electron-photon interaction Hamiltonian, see e.g.~\cite{Weisskopf,Berestetskii}.

The abundance of energy states in many-electron systems and the inherent weakness of 
the $\mathcal{O}(\alpha_S^2)$ scattering signal imply that photon scattering in materials is
always dominated by nearly resonant transitions through intermediate virtual states.
This explains why the Kramers-Heisenberg approximation is ubiquitous in spectroscopy 
with synchrotron 
radiation \cite{Ma1,Ma2,Gelmukhanov1,Eisebitt,Shirley,Ament,Glatzel,Gelmukhanov2}.
The Kramers-Heisenberg formula has been successfully applied e.g.~to $\mathrm{CaF}_2$ \cite{deGroot},
lanthanum and lanthanum compounds \cite{Moewes1b,Moewes4,Taguchi} as well as compounds
of other rare-earth elements \cite{Moewes1,Moewes2,Moewes3,Hunt},
titanium and titanium compounds \cite{Jimenez}, cobalt compounds \cite{Magnuson,Wang},
lithium fluoride \cite{Kikas}, silicon and aluminum and their 
compounds \cite{Szlachetko1,Szlachetko2,Zhang}, zinc oxide \cite{Preston},
aequous solutions of transition metals \cite{Bokarev,Green}, and $\mathrm{N}_2$ \cite{Kjellsson}.

The second resonance condition (\ref{eq:res2}) cannot be fulfilled if $|i\rangle$ is the ground 
state $|g\rangle$ of the scattering system, or if $E_i-E_g<\hbar ck'$. In these cases only the first
term in Eq.~(\ref{eq:KHdd}) (which is known as the ``resonant term'') is kept, while the second 
term (often denoted as the ``non-resonant'' or ``anti-resonant'' term) can be discarded.

We will display the corresponding scattering cross sections with continuous external electron 
states $|i\rangle$ or $|f\rangle$ only in the Kramers-Heisenberg approximation. The 
corresponding $\mathcal{O}(\alpha_S^2)$ correct formulae like (\ref{eq:master1}) can be inferred from 
the corresponding Kramers-Heisenberg formulae through reversing the steps that led from (\ref{eq:master1}) 
to (\ref{eq:KHdd}).

\subsection{Differential scattering cross section for $\bm{|i\rangle}$ discrete and $\bm{|f\rangle}$
  continuous}

\begin{figure}[htb]
\scalebox{0.9}{\includegraphics{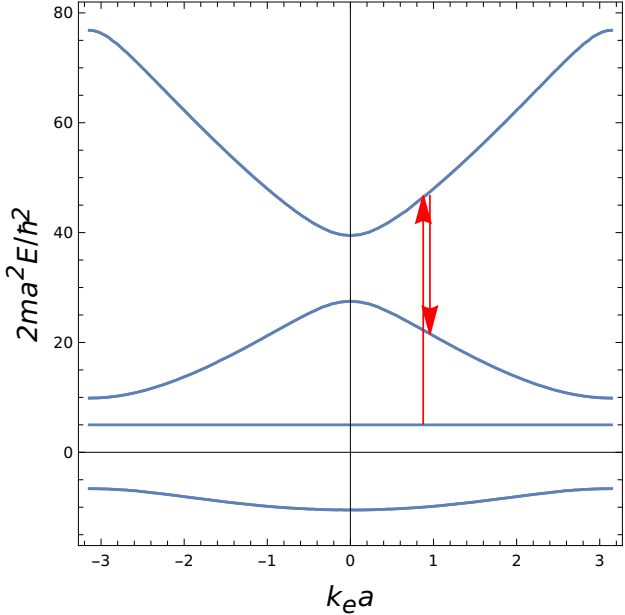}}
\caption{\label{fig:dcscatter}
  Scattering from a $\bm{k}_e$-independent discrete atomic energy level
  into an energy band.
  In this case the scattering is resonantly enhanced through
  an intermediate band state.
The depicted bands correspond to the lowest three energy
bands in a Kronig-Penney model $V(x)=\mathcal{W}\sum_n\delta(x-na)$
with parameter $m\mathcal{W}a/\hbar^2=-\,3$.}
\end{figure}

This case applies e.g.~to excitation of an electron from a donor level into the conduction band of a 
semiconductor if the energy absorption does not occur as a consequence of direct photon absorption
(as described in Sec.~\ref{subsec:ad2c}), but through photon scattering. We formulate the corresponding
Kramers-Heisenberg formula for the case that the final states reside in energy bands $E_n(\bm{k}_{e})$,
$|f\rangle=|n,\bm{k}_{e}\rangle$.
A schematic involving resonantly enhanced
scattering through intermediate energy band states is depicted in Fig.~\ref{fig:dcscatter}

Transitions can involve intermediate and final Bloch states with
arbitrary wave vector $\bm{k}_e$, because the limit $\lambda_e\ge 2a$ on Bloch wavelengths 
implies that the plane wave factors vary slowly over the extent of the atomic wave functions.
The emitting interband transition is a direct transition because the emitted photon wavelength
satisfies $\lambda\gg 2a$ and therefore $k\ll\pi/a$, see Eq.~(\ref{eq:direct}).

The differential photon 
scattering cross section into $d\Omega dk'$ for the scattered photons implies an integration over the 
energy surface $E_f\equiv E_n(\bm{k}_{e})=E_i+\hbar ck-\hbar ck'$ in the conduction band,
\begin{eqnarray}\nonumber
\frac{d\sigma}{d\Omega dk'}&=&\alpha_S^2\hbar c^3 kk'^3
\int_{E_n(\bm{k}_{e})=E_i+\hbar ck-\hbar ck'}
\frac{d^2\bm{k}_{e\|}}{|\partial E_n(\bm{k}_{e})/\partial\bm{k}_{e}|}
\\ \nonumber
&&\times\!\left|\sum_v
\frac{\langle n,\bm{k}_{e}|\bm{\epsilon}'_\gamma\cdot\op{x}|v\rangle
\langle v|\bm{\epsilon}_\gamma\cdot\op{x}|i\rangle}{
\omega_{vi}-ck-\mathrm{i}\eta}
\right.
\\ \label{eq:KHdc}
&&+\!\left.\sum_v
\frac{\langle n,\bm{k}_{e}|\bm{\epsilon}_\gamma\cdot\op{x}|v\rangle
\langle v|\bm{\epsilon}'_\gamma\cdot\op{x}|i\rangle}{
\omega_{vi}+ck'-\mathrm{i}\eta}\right|^2\!\!.
\end{eqnarray}

In this case, the assumption of approximately constant Kramers-Heisenberg dispersion factor
across the constant energy surface
relates the differential scattering cross section to the density of states (\ref{eq:rhonE}),
\begin{eqnarray}\nonumber
\frac{d\sigma}{d\Omega dk'}&=&\alpha_S^2\hbar c^3 kk'^3
\varrho_{n,V}(E_i+\hbar ck-\hbar ck')
\\ \nonumber
&&\times\!
\left|\sum_v
\frac{\langle u_n(\bm{k}_{e})|\bm{\epsilon}'_\gamma\cdot\op{x}|v\rangle
\langle v|\bm{\epsilon}_\gamma\cdot\op{x}|i\rangle}{
\omega_{vi}-ck-\mathrm{i}\eta}
\right.
\\ \label{eq:KHdcrho}
&&+\!\left.\sum_v
\frac{\langle u_n(\bm{k}_{e})|\bm{\epsilon}_\gamma\cdot\op{x}|v\rangle
\langle v|\bm{\epsilon}'_\gamma\cdot\op{x}|i\rangle}{
\omega_{vi}+ck'-\mathrm{i}\eta}\right|^2\!\!.
\end{eqnarray}

Just as for the dipole factors in the absorption cross 
sections (\ref{eq:absB3},\ref{eq:absB4},\ref{eq:absB2b},\ref{eq:absB2c}), 
the assumption of approximately constant dispersion factor across the constant
energy surface $E_n(\bm{k}_{e})=E_i+\hbar ck-\hbar ck'$ could be justified if that energy surface
is small compared to the typical area dimension $4\pi^2/V^{2/3}$ of the Brillouin zone.
The integral in (\ref{eq:KHdc}) then spans a relatively small surface area in the Brillouin zone 
such that the intermediate and final Bloch wave functions do not vary a lot with $\bm{k}_e$.

 For the contributions from intermediate Bloch states 
\begin{equation}
|n',\bm{k}'_e\rangle=\sqrt{\frac{V}{8\pi^3}}
\exp(\mathrm{i}\bm{k}'_e\cdot\op{x})|u_{n'}(\bm{k}'_e)\rangle
\end{equation}
to the sum over virtual intermediate states $|v\rangle$ in the dispersion factors, 
we note with (\ref{eq:sumR}) that
\begin{eqnarray}\nonumber
&&\sum_{n'}\int\!d^3\bm{k}'_e\,
\langle n,\bm{k}_{e}|\bm{\epsilon}_\gamma\cdot\op{p}|n',\bm{k}'_e\rangle\langle n',\bm{k}'_{e}|\ldots
\\ \nonumber
&&=\sum_{n'}
\langle u_n(\bm{k}_{e})|\bm{\epsilon}_\gamma\cdot(\op{p}+\hbar\bm{k}_e)|u_{n'}(\bm{k}_e)\rangle_V
\langle n',\bm{k}_{e}|\ldots
\\ \nonumber
&&=\mathrm{i}\frac{m}{\hbar}\sum_{n'}
\langle u_n(\bm{k}_{e})|\bm{\epsilon}_\gamma\cdot[H_0(\bm{k}_e),\op{x}]|u_{n'}(\bm{k}_e)\rangle_V
\\ \label{eq:interu}
&&\quad\times
\langle n',\bm{k}_{e}|\ldots,
\end{eqnarray}
where
\begin{equation}\label{eq:H0u}
H_0(\bm{k}_e)=\frac{1}{2m}(\op{p}+\hbar\bm{k}_e)^2+V(\op{x}) 
\end{equation}
is the lattice Hamiltonian for the periodic Bloch factors $\langle\bm{x}|u_{n}(\bm{k}_e)\rangle$.
This implies that the contribution from the intermediate virtual band states to the scattering cross
section amounts to a summation over band states with the same electron wave vector $\bm{k}_e$.
 Furthermore, the matrix element of the Bloch states reduces to an integral over the
Wigner-Seitz cell, where the Bloch wave functions $\langle\bm{x}|n,\bm{k}_e\rangle$ are replaced 
with the corresponding Bloch factors $\langle\bm{x}|u_{n}(\bm{k}_e)\rangle$.

\subsection{Differential scattering cross section for $\bm{|i\rangle}$ continuous
  and $\bm{|f\rangle}$ discrete}

\begin{figure}[htb]
\scalebox{0.9}{\includegraphics{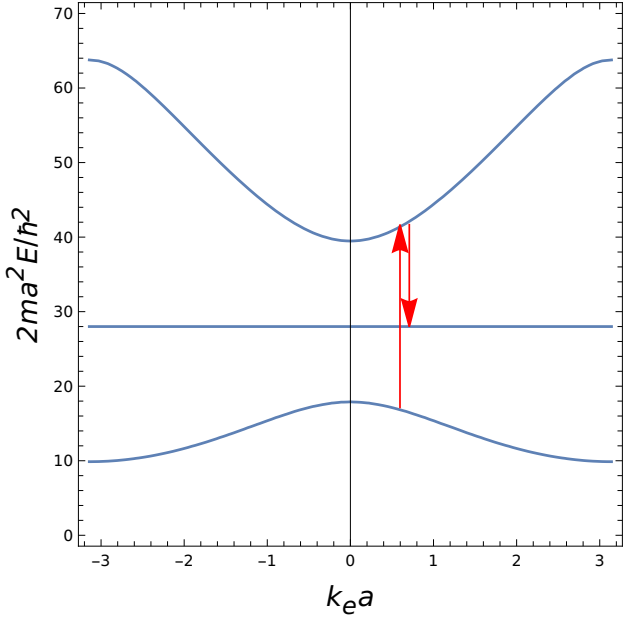}}
\caption{\label{fig:cdscatter}
  Scattering into a $\bm{k}_e$-independent discrete atomic energy level.
  In this case the scattering is resonantly enhanced through
  an intermediate band state.
The depicted bands correspond to the second and third energy
band in a Kronig-Penney model $V(x)=\mathcal{W}\sum_n\delta(x-na)$
with parameter $m\mathcal{W}a/\hbar^2=-\,7$.}
\end{figure}

This case would apply e.g.~to a case where an electron is promoted from a valence band state into an
acceptor state through photon scattering instead of straight photon absorption. We therefore use 
valence band states as initial states, $|i\rangle=|n,\bm{k}_{e}\rangle$.
A schematic involving resonantly enhanced
scattering through intermediate energy band states is depicted in Fig.~\ref{fig:cdscatter}

Transitions can involve initial and intermediate Bloch states with
arbitrary wave vector $\bm{k}_e$, because the limit $\lambda_e\ge 2a$ on Bloch wavelengths 
implies that the plane wave factors vary slowly over the extent of the atomic wave functions.
The absorbing interband transition is a direct transition because the emitted photon wavelength
satisfies $\lambda\gg 2a$ and therefore $k\ll\pi/a$, see Eq.~(\ref{eq:direct}).

The differential photon 
scattering cross section into $d\Omega dk'$ for the scattered photons implies an integration over 
the energy surface $E_i\equiv E_n(\bm{k}_{e})=E_f+\hbar ck'-\hbar ck$ in the valence band,
\begin{eqnarray}\nonumber
\frac{d\sigma}{d\Omega dk'}&=&\alpha_S^2\hbar c^3 kk'^3
\int_{E_n(\bm{k}_{e})=E_f+\hbar ck'-\hbar ck}
\frac{d^2\bm{k}_{e\|}}{|\partial E_n(\bm{k}_{e})/\partial\bm{k}_{e}|}
\\ \nonumber
&&\times\!\left|\sum_v
\frac{\langle f|\bm{\epsilon}'_\gamma\cdot\op{x}|v\rangle
\langle v|\bm{\epsilon}_\gamma\cdot\op{x}| n,\bm{k}_{e}\rangle}{
\omega_{vi}-ck-\mathrm{i}\eta}
\right.
\\ \label{eq:KHcd}
&&+\!\left.\sum_v
\frac{\langle f|\bm{\epsilon}_\gamma\cdot\op{x}|v\rangle
\langle v|\bm{\epsilon}'_\gamma\cdot\op{x}| n,\bm{k}_{e}\rangle}{
\omega_{vi}+ck'-\mathrm{i}\eta}\right|^2\!\!,
\end{eqnarray}
and the assumption of approximately constant dispersion factor relates this again to
the density of states (\ref{eq:rhonE}),
\begin{eqnarray}\nonumber
\frac{d\sigma}{d\Omega dk'}&=&\alpha_S^2\hbar c^3 kk'^3
\varrho_{n,V}(E_f+\hbar ck'-\hbar ck)
\\ \nonumber
&&\times\!
\left|\sum_v
\frac{\langle f|\bm{\epsilon}'_\gamma\cdot\op{x}|v\rangle
\langle v|\bm{\epsilon}_\gamma\cdot\op{x}| u_n(\bm{k}_{e})\rangle}{
\omega_{vi}-ck-\mathrm{i}\eta}
\right.
\\ \label{eq:KHcdrho}
&&+\!\left.\sum_v
\frac{\langle f|\bm{\epsilon}_\gamma\cdot\op{x}|v\rangle
\langle v|\bm{\epsilon}'_\gamma\cdot\op{x}| u_n(\bm{k}_{e})\rangle}{
\omega_{vi}+ck'-\mathrm{i}\eta}\right|^2\!\!.
\end{eqnarray}

The observations (\ref{eq:interu},\ref{eq:H0u}) concerning the contributions from intermediate
virtual band states also apply here.

\subsection{Differential scattering cross section for both $\bm{|i\rangle}$
  and $\bm{|f\rangle}$ continuous}

This applies e.g.~to photon scattering off free electrons. The resonance 
conditions (\ref{eq:res1}) or (\ref{eq:res2}) cannot be fulfilled in this case
and the elastic Thomson scattering term (the first term in Eq.~(\ref{eq:master1p}))
dominates low-energy photon scattering off free electrons. The factor $\delta_{fi}$
in the discrete-to-discrete Thomson term in (\ref{eq:master1p}) is replaced with
$\int d^3\bm{k}'_e\delta(\bm{k}'_e-\bm{k}_e)\delta(\bm{0})8\pi^3/\mathcal{V}=1$ for
scattering of free electrons with initial momentum $\hbar\bm{k}_e$ into outgoing
momentum eigenstates $|f\rangle=|\bm{k}'_e\rangle$. This yields the well-known result
\begin{equation}
\frac{d\sigma}{d\Omega}
=\left(\frac{\alpha_S\hbar}{mc}\right)^2
\left(\bm{\epsilon}'_\gamma\cdot\bm{\epsilon}_\gamma\right)^2.
\end{equation}
The Kramers-Heisenberg dispersion terms provide corrections to the Thomson term which in leading 
order scale like $\mathcal{O}(\hbar k/mc)$. However, the quantum optics result (\ref{eq:master1p})
is not useful for calculating corrections to Thomson scattering off free electrons 
because Compton scattering is dealt with by the Klein-Nishina formula.

\begin{figure}[htb]
\scalebox{0.9}{\includegraphics{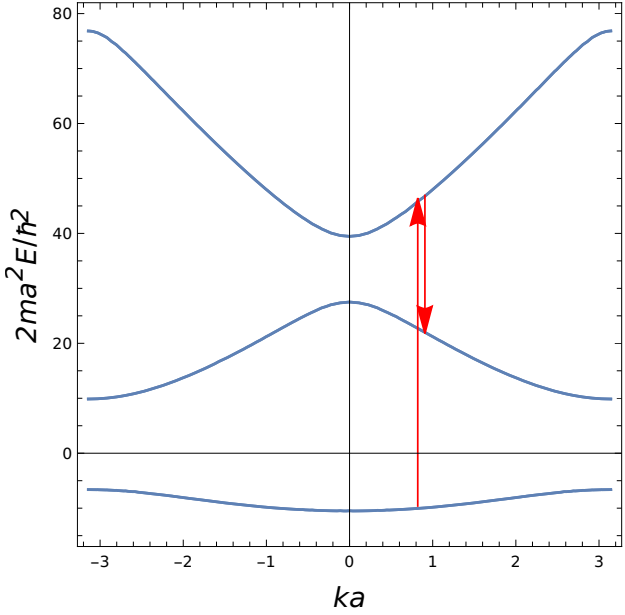}}
\caption{\label{fig:ccscatter}
  Scattering between energy bands.
  In this case the scattering is resonantly enhanced through
  an intermediate band state.
The depicted bands correspond to the lowest three energy
bands in a Kronig-Penney model $V(x)=\mathcal{W}\sum_n\delta(x-na)$
with parameter $m\mathcal{W}a/\hbar^2=-\,3$.
The interband transitions are direct transitions because the photon wave vectors
satisfy $k\ll\pi/a$.}
\end{figure}

On the other hand, scattering between continuous initial and final electron states also applies
to scattering between energy bands $|i\rangle=|n,\bm{k}_e\rangle\to|f\rangle=|n',\bm{k}_e\rangle$, 
where dipole approximation yields again direct interband transitions $\bm{k}'_e=\bm{k}_e$.
Adapting the Kramers-Heisenberg relation to this situation became important with
the availability of synchrotron light sources for inelastic X-ray scattering
between energy bands in materials \cite{Ma1,Ma2,Eisebitt,Shirley}.
A schematic involving resonantly enhanced
scattering through intermediate energy band states is depicted in Fig.~\ref{fig:ccscatter}

The scattering cross section in the Kramers-Heisenberg approximation with 
scattering through intermediate band states $|v,\bm{k}''_e\rangle$ is
\begin{eqnarray}\label{eq:KHcc1}
&&\frac{d\sigma}{d\Omega}=\alpha_S^2 c^2 kk'^3
\\ \nonumber
&&\times\!\left|\sum_{v}
\frac{\langle u_{n'}(\bm{k}_e)|\bm{\epsilon}'_\gamma\cdot\op{x}|u_{v}(\bm{k}_e)\rangle_V
\langle u_{v}(\bm{k}_e)|\bm{\epsilon}_\gamma\cdot\op{x}|u_{n}(\bm{k}_e)\rangle_V}{
\omega_{v,n}(\bm{k}_e)-ck-\mathrm{i}\eta}
\right.
\\ \nonumber
&&+\!\left.\sum_{v}
\frac{\langle u_{n'}(\bm{k}_e)|\bm{\epsilon}_\gamma\cdot\op{x}|u_{v}(\bm{k}_e)\rangle_V
\langle u_{v}(\bm{k}_e)|\bm{\epsilon}'_\gamma\cdot\op{x}|u_{n}(\bm{k}_e)\rangle_V}{
\omega_{v,n}(\bm{k}_e)+ck'-\mathrm{i}\eta}\right|^2\!\!,
\end{eqnarray}
where $ck'=ck-\omega_{n',n}(\bm{k}_e)$ and Eqs.~(\ref{eq:sumR},\ref{eq:interu}) imply
that all transitions occur at the same point $\bm{k}_e$ in the Brillouin zone.

The differential scattering cross section per lattice cell for 
transitions from a valence band $E_n(\bm{k}_e)$ into a conduction 
band $E_{n'}(\bm{k}_e)$ is
\begin{eqnarray}\label{eq:KHcc2}
&&\frac{d\sigma}{d\Omega dk'}=\alpha_S^2\hbar c^3 kk'^3
\frac{V}{8\pi^3}
\\ \nonumber
&&\times\int_{E_{n'}(\bm{k}_e)-E_n(\bm{k}_e)=\hbar c(k-k')}
\frac{d^2\bm{k}_{e\|}}{|\partial[E_{n'}(\bm{k}_e)-E_n(\bm{k}_e)]/\partial\bm{k}_e|}
\\ \nonumber
&&\times\!\left|\sum_{v}
\frac{\langle u_{n'}(\bm{k}_e)|\bm{\epsilon}'_\gamma\cdot\op{x}|u_{v}(\bm{k}_e)\rangle_V
\langle u_{v}(\bm{k}_e)|\bm{\epsilon}_\gamma\cdot\op{x}|u_{n}(\bm{k}_e)\rangle_V}{
\omega_{v,n}(\bm{k}_e)-ck-\mathrm{i}\eta}
\right.
\\ \nonumber
&&+\!\left.\sum_{v}
\frac{\langle u_{n'}(\bm{k}_e)|\bm{\epsilon}_\gamma\cdot\op{x}|u_{v}(\bm{k}_e)\rangle_V
\langle u_{v}(\bm{k}_e)|\bm{\epsilon}'_\gamma\cdot\op{x}|u_{n}(\bm{k}_e)\rangle_V}{
\omega_{v,n}(\bm{k}_e)+ck'-\mathrm{i}\eta}\right|^2\!\!.
\end{eqnarray}

We can express this through the joint density of states (\ref{eq:jointrho}) 
at the energy transfer $\Delta E=\hbar c(k-k')$ if 
the Kramers-Heisenberg dispersion factor is approximately constant
over the constant energy surface $E_{n'}(\bm{k}_e)-E_n(\bm{k}_e)=\hbar c(k-k')$,
\begin{eqnarray}\label{eq:KHcc3}
&&\frac{d\sigma}{d\Omega dk'}=\alpha_S^2\hbar c^3 kk'^3
\varrho_{n'n,V}(\hbar ck-\hbar ck')
\\ \nonumber
&&\times\!\left|\sum_{v}
\frac{\langle u_{n'}(\bm{k}_e)|\bm{\epsilon}'_\gamma\cdot\op{x}|u_{v}(\bm{k}_e)\rangle_V
\langle u_{v}(\bm{k}_e)|\bm{\epsilon}_\gamma\cdot\op{x}|u_{n}(\bm{k}_e)\rangle_V}{
\omega_{v,n}(\bm{k}_e)-ck-\mathrm{i}\eta}
\right.
\\ \nonumber
&&+\!\left.\sum_{v}
\frac{\langle u_{n'}(\bm{k}_e)|\bm{\epsilon}_\gamma\cdot\op{x}|u_{v}(\bm{k}_e)\rangle_V
\langle u_{v}(\bm{k}_e)|\bm{\epsilon}'_\gamma\cdot\op{x}|u_{n}(\bm{k}_e)\rangle_V}{
\omega_{v,n}(\bm{k}_e)+ck'-\mathrm{i}\eta}\right|^2\!\!.
\end{eqnarray}

\section{Conclusions}
\label{sec:conc}

The leading order equations for photon emission have been summarized
in Sec.~\ref{sec:em}, for photon absorption in Sec.~\ref{sec:abs}, and for
photon scattering in Sec.~\ref{sec:sc}. In each section, the different cases
of discrete-to-discrete, discrete-to-continuous, continuous-to-discrete
and continuous-to-continuous electronic transitions have been described
in separate subsections. The different cases of photon transitions differ
in terms of the quantity that is calculated as a measure for the transition:
emission rates for photon emission, absorption cross sections for photon absorption,
and scattering cross sections for photon scattering. The different cases of
electronic transitions differ in the appropriate factors for incoming and outgoing
electronic states in terms of densities of states or joint densities of states,
or through integrations over $\bm{k}_e$-space,
if continuous states are involved. This compendium can hopefully serve as a
concise manual for the many researchers who navigate the 
landscape of radiative transition equations for their work in
spectroscopy, quantum optics, or photonics.



\appendix
\section{Continuous states}
\label{sec:cont}

Ionization of materials and ion-electron recombination involve states in the continuous energy spectrum
of a material. Transitions in materials which involve energy bands also involve states in an energy 
continuum. Equations for transitions involving continuous states, e.g.~the Golden Rule, are 
often expressed in terms of densities of states in the energy scale. However, transition 
probabilities between energy bands in condensed materials are more commonly derived in terms of the
quasiperiodic Bloch energy eigenstates $|n,\bm{k}_e\rangle$,
\begin{eqnarray}\nonumber
\langle\bm{x}|n,\bm{k}_e\rangle&=&\psi_n(\bm{k}_e,\bm{x})
\\ \label{eq:blochpsi}
&=&\sqrt{\frac{V}{8\pi^3}}\exp(\mathrm{i}\bm{k}_e\cdot\bm{x})u_n(\bm{k}_e,\bm{x}),
\end{eqnarray}
where $V$ is the volume of the Wigner-Seitz cell and $u_n(\bm{k}_e,\bm{x})$ are the
periodic Bloch factors for the energy band $E_n(\bm{k}_e)$. The Bloch energy
eigenfunctions $\psi_n(\bm{k}_e,\bm{x})$
are periodically perturbed plane waves with normalization
\begin{equation}\label{eq:normbloch1}
  \langle n',\bm{k}'_e|n,\bm{k}_e\rangle=\delta_{nn'}\delta(\bm{k}_e-\bm{k}'_e).
  \end{equation}

We extract the factor $\sqrt{V/8\pi^3}$ from the Bloch factors, because with this definition
the property (\ref{eq:normbloch1}) of the Bloch energy eigenfunctions
implies normalization of the periodic
Bloch factors $\langle\bm{x}|u_n(\bm{k}_e)\rangle=u_n(\bm{k}_e,\bm{x})$ to the Wigner-Seitz cell,
\begin{eqnarray}\nonumber
  \langle n',\bm{k}_e|n,\bm{k}_e\rangle_V&\equiv&\int_V\!d^3\bm{x}\,u^+_{n'}(\bm{k}_e,\bm{x})
  u_n(\bm{k}_e,\bm{x})
  \\
  &=&\delta_{nn'},
\end{eqnarray}
where the integration is over the Wigner-Seitz cell and both Bloch factors must refer to the same
wave vector $\bm{k}_e$ in the Brillouin zone.

The connections between densities of states in the energy scale and wave vector parametrizations for
continuous states are encoded
in the completeness relations, which involve sums over the discrete energy 
eigenvalues $E_j$ and the continuous parts $C$ of the spectrum,
\begin{eqnarray}\nonumber
1&=&\sum_{j,\nu}|E_j,\nu\rangle\langle E_j,\nu|
\\ \label{eq:completeE}
&&+\sum_{\nu}\int_{C}\!dE\,|E,\nu\rangle\tilde{\varrho}(E,\nu)\langle E,\nu|
\\ \nonumber
&=&\sum_{j,\nu}|E_j,\nu\rangle\langle E_j,\nu|
\\ \label{eq:completeu}
&&+\sum_{n}\int_{\mathcal{B}}\!d^3\bm{k}_e\,|n,\bm{k}_e\rangle\langle n,\bm{k}_e|.
\end{eqnarray}
The integral $\int_{\mathcal{B}}\!d^3\bm{k}_e\ldots$ covers the first Brillouin zone.
The energy integral $\int_{C}\!dE\ldots$ covers all the continuous energy eigenvalues,
i.e.~in a condensed material the integration domain $C=\sum_n\oplus\, C_n$ 
covers all the energy bands.
The sum $\sum_{\nu}$ over degeneracy indices for the energy eigenstates $|E,\nu\rangle$
in the continuous part $C$ of the spectrum can involve summations over discrete quantum
numbers or integrations over continuous quantum numbers. 

The contribution from the energy band $E_n(\bm{k}_e)$ to the density of states follows from
\begin{equation}\label{eq:trans0}
d^3\bm{k}_e=d^2\bm{k}_{e\|}dk_{e,\perp}\to d^2\bm{k}_{e\|}
\frac{dE}{|\partial E_n(\bm{k}_e)/\partial\bm{k}_e|},
\end{equation}
where $d^2\bm{k}_{e\|}$ denotes an integration measure on the surface of constant 
energy $E_n(\bm{k}_e)$. This implies
\begin{eqnarray}\nonumber
&&\int_{\mathcal{B}}\!d^3\bm{k}_e\,|n,\bm{k}_e\rangle\langle n,\bm{k}_e|
\\ \label{eq:trans1}
&&\to\int_{C_n}\!dE\int_{E_n(\bm{k}_e)=E}\!d^2\bm{k}_{e\|}\,
\frac{|n,\bm{k}_e\rangle\langle n,\bm{k}_e|}{|\partial E_n(\bm{k}_e)/\partial\bm{k}_e|}.
\end{eqnarray}

If we agree to use the continuous variables included in the integration measure $d^2\bm{k}_{e\|}$
as degeneracy indices $\nu$, $\sum_\nu\to \int\!d^2\bm{k}_{e\|}$,
and set $|E,\bm{k}_\|\rangle=|n,\bm{k}_e\rangle$ for $E=E_n(\bm{k}_e)$, comparison of
(\ref{eq:trans1}) with (\ref{eq:completeE}) tells us that the contribution from the 
energy band $E_n(\bm{k})$ to the partial density of
states $\tilde{\varrho}(E,\nu)=\sum_n\tilde{\varrho}_n(E,\nu)$ is
\begin{equation}\label{eq:partrho1}
 \tilde{\varrho}_n(E,\nu)\equiv\tilde{\varrho}_n(E,\bm{k}_\|)=
\frac{1}{|\partial E_n(\bm{k}_e)/\partial\bm{k}_e|}.
\end{equation} 

The normalization of $\tilde{\varrho}(E,\nu)$ depends on the normalization of the continuous states
$|E,\nu\rangle$. The expression (\ref{eq:varrhoEx}) for energies in the continuous
part of the spectrum,
\begin{equation}\label{eq:LDOS}
\varrho(E,\bm{x})dE=
\sum_\nu\langle\bm{x}|E,\nu\rangle\tilde{\varrho}(E,\nu)\langle E,\nu|\bm{x}\rangle dE
\end{equation}
gives the local density of continuous states per volume $V$ at location $\bm{x}$ and 
with energies in the interval $[E,E+dE]$, i.e.~scaling the states $|E,\nu\rangle$ with 
a factor $\xi$ scales $\tilde{\varrho}(E,\nu)$ by $|\xi|^{-2}$. 

We can switch from the \textit{a priori}
continuous degeneracy indices $\nu_c$, which provide coordinates on the constant energy surface,
to discrete indices $\nu_d$ through harmonic analysis,
\begin{equation*}
\int\!d^2\nu_c\,|E,\nu_c\rangle\tilde{\varrho}(E,\nu_c)\langle E,\nu_c|
=\sum_{\nu_d}|E,\nu_d\rangle\tilde{\varrho}(E,\nu_d)\langle E,\nu_d|.
\end{equation*}
Both parametrizations will yield the same local density of states $\varrho(E,\bm{x})$. 
The local density of states (per spin state) for nonrelativistic free
electrons with plane wave states $|\bm{k}_e\rangle$ is
\begin{eqnarray}\nonumber
\varrho_e(E,\bm{x})&=&
\int_{E(\bm{k}_e)=E}\!d^2\bm{k}_{e\|}\frac{|\langle\bm{x}|\bm{k}_e\rangle|^2
}{|\partial E(\bm{k}_e)/\partial\bm{k}_e|}
\\ \nonumber
&=&\frac{1}{8\pi^3}
\int_{E(\bm{k}_e)=E}\frac{d^2\bm{k}_{e\|}}{|\partial E(\bm{k}_e)/\partial\bm{k}_e|}
\\ \label{eq:ldos1}
&=&\frac{\Theta(E)}{2\pi^2\hbar^3}\sqrt{2m^3E},
\end{eqnarray}
and this is independent of position. However, the local density of electron states 
(per spin state) in the
energy band $E_n(\bm{k}_e)$ involves the Bloch factors,
\begin{equation}\label{eq:ldos2}
\varrho_n(E,\bm{x})=\frac{V}{8\pi^3}
\int_{E_n(\bm{k}_e)=E}\!d^2\bm{k}_{e\|}\frac{|u_n(\bm{k}_e,\bm{x})|^2
}{|\partial E(\bm{k}_e)/\partial\bm{k}_e|},
\end{equation}
where $V$ is the volume of the Wigner-Seitz cell. The density of states (\ref{eq:rhonE}) 
emerges from the local density of states (\ref{eq:ldos2}) after integration over 
a Wigner-Seitz cell,
\begin{equation}\label{eq:ldos3}
\varrho_{n,V}(E)=\int_V\!d^3\bm{x}\,\varrho_n(E,\bm{x}).
\end{equation}

\section{The quantum fields in the quantum optics Hamiltonian}

The electron field $\Psi_s(\bm{x},t)$ and the photon field $\bm{A}(\bm{x},t)$
in the Coulomb gauge Hamiltonian (\ref{eq:HQO}-\ref{eq:Hegamma}) are 
quantum fields in the Heisenberg picture which are related to
the time-independent quantum fields of the Schr\"odinger picture through
\begin{equation}\label{eq:PsiH}
\Psi_s(\bm{x},t)=\exp(\mathrm{i}Ht/\hbar)\Psi_s(\bm{x})\exp(-\,\mathrm{i}Ht/\hbar),
\end{equation} 
\begin{equation}\label{eq:AH}
\bm{A}(\bm{x},t)=\exp(\mathrm{i}Ht/\hbar)\bm{A}(\bm{x})\exp(-\,\mathrm{i}Ht/\hbar).
\end{equation}

Canonical quantization implies the anticommutation relations for electron operators
\begin{equation*}
  \{\Psi_s(\bm{x}),\Psi_{s'}^+(\bm{x}')\}=\delta(\bm{x}-\bm{x}'),\quad
  \{\Psi_s(\bm{x}),\Psi_{s'}(\bm{x}')\}=0,
\end{equation*}
and commutation relations for photon operators
(with $E_{i}(\bm{x})=-\,\dot{A}_{i}(\bm{x},t)|_{t\to 0}$)
\begin{equation}\label{eq:commA1}
[A_{i}(\bm{x}),E_{j}(\bm{x}')]
=-\,\frac{\mathrm{i}\hbar}{\epsilon_0}\delta_{ij}^\perp(\bm{x}-\bm{x}'),
\end{equation}
\begin{equation}\label{eq:commA2}
[A_{i}(\bm{x}),A_{j}(\bm{x}')]=0,\quad [E_{i}(\bm{x}),E_{j}(\bm{x}')]=0,
\end{equation}
where
\begin{equation}
  \delta_{ij}^\perp(\bm{x})
  =\frac{1}{(2\pi)^3}
\int\!d^3\bm{k}\left(\delta_{ij}
-\frac{k_i k_j}{\bm{k}^2}\right)
\exp(\mathrm{i}\bm{k}\cdot\bm{x})
\end{equation}
is the transverse $\delta$ function.

The mode expansion of $A_{i}(\bm{x})$ contains the photon operators $a^+_\alpha(\bm{k})$
which create photons with momentum $\hbar\bm{k}$ and polarization $\bm{\epsilon}_\alpha(\bm{k})$,
\begin{eqnarray}\nonumber
  \bm{A}(\bm{x})&=&\sqrt{\frac{\hbar\mu_0 c}{8\pi^3}}
\int\!\frac{d^3\bm{k}}{\sqrt{2k}}\sum_{\alpha=1}^2
\bm{\epsilon}_\alpha(\bm{k})\Big(
a_\alpha(\bm{k})
\exp(\mathrm{i}\bm{k}\cdot\bm{x})
\\ \label{eq:Amode}
&&+\,a_{\alpha}^+(\bm{k})
\exp(-\,\mathrm{i}\bm{k}\cdot\bm{x})\Big).
\end{eqnarray}


The electron creation operators $\Psi^+_s(\bm{x})$ create e.g.~single-electron states
with spinor components $\psi_s(\bm{x},t)$ through
\begin{equation}\label{eq:1e}
\bm{|}\psi(t)\bm{\rangle}=\sum_s\int\!d^3\bm{x}\,\Psi^+_s(\bm{x})\bm{|}0\bm{\rangle}\psi_s(\bm{x},t),
\end{equation}
or general many-electron states,
\begin{eqnarray}\nonumber
  \bm{|}\psi(t)\bm{\rangle}&=&\sum_{s,s',\ldots}\int\!d^3\bm{x}\int\!d^3\bm{x}'
  \,\Psi^+_s(\bm{x})\Psi^+_{s'}(\bm{x}')\ldots\bm{|}0\bm{\rangle}
  \\ \label{eq:Ne}
  &&\times\psi_{s,s',\ldots}(\bm{x},\bm{x}',\ldots,t).
\end{eqnarray}
The many-electron states (\ref{eq:Ne}) are usually approximated through products
of orthonormalized single-electron states,
\begin{eqnarray}\nonumber
  &&\psi_{s_1,s_2,\ldots,s_N}(\bm{x}_1,\bm{x}_2,\ldots,\bm{x}_N,t)
  \to\psi_{n_1,s_1}(\bm{x}_1,t)
  \\ \label{eq:Ne2}
  &&\times\psi_{n_2,s_2}(\bm{x}_2,t)\ldots\psi_{n_N,s_N}(\bm{x}_N,t).
  \end{eqnarray}

The many-particle state in a material can then be thought of as the product
of a many-electron state of the form (\ref{eq:Ne},\ref{eq:Ne2}) with corresponding
many-particle states for the pertinent nuclei.
The expectation value of the kinetic electron energy operator
$\mathcal{H}_e$ (\ref{eq:HQOa}) for the many-particle state of the material
then generates the sum of
the kinetic energy densities of all the fundamental nonrelativistic electrons,
and the expectation value of the potential operator for the electrons,
i.e.~the sum of relevant terms from (\ref{eq:HQOab}),
\begin{eqnarray}\nonumber \label{eq:Ve}
  \mathcal{V}_e(\bm{x})&=&\int\!d^3\bm{x}'\,\mathcal{H}_{ee}(\bm{x},\bm{x}')
    \\
    &&+\,2\sum_{b=\mathrm{nuclei}}\int\!d^3\bm{x}'\,\mathcal{H}_{eb}(\bm{x},\bm{x}'),
  \end{eqnarray}
generates an effective single-electron potential.
We use the Schr\"odinger picture in (\ref{eq:Ve}), such that the wave functions
in our many-particle quantum states are time-dependent but the operators are
time-independent, see (\ref{eq:Ne}).

The Coulomb terms in (\ref{eq:HQOab}) automatically generate the
exchange interaction terms in $\bm{\langle}\mathcal{V}_e(\bm{x})\bm{\rangle}$
through Fermi statistics, but for realistic potentials
in materials we would also have to include first order relativistic corrections
like spin-orbit coupling. Furthermore, instead of summation over bare nuclei, we may
decide to allocate core electrons to the nuclei and not count them separately towards
the electronic states in the material, in the interest of efficiency. However, these details
do not impact the spectroscopic formulae in this paper.
The important observation is that even in a complex material we are still
considering fundamental electrons which are moving in a potential, e.g.~a periodic
lattice potential
in a solid material, and that couple to photons through the minimal coupling 
terms $\mathcal{H}_{e\gamma}$ (\ref{eq:Hegamma}) where $m=m_e$ is the electron mass. 
Furthermore, these electrons are described on the quantum mechanical level by the
Hamiltonian (\ref{eq:Hfirstq}) with an effective potential $V(\op{x})$, such that
(\ref{eq:lengthform}) still holds. Eq.~(\ref{eq:Hfirstq}) and the transition
between velocity and length form
would not hold anymore if we have to take into account relativistic corrections to the
kinetic energy, e.g.~for deep core electrons in materials with heavy atoms.

The representation of the $\bm{x}$-space electron operators in terms of free $\bm{k}_e$-space
electron operators is
\begin{equation}\label{eq:Psi2afree}
  \Psi_s(\bm{x})=\frac{1}{\sqrt{2\pi}^3}\int\!d^3\bm{k}_e\,\exp(\mathrm{i}\bm{k}_e\cdot\bm{x})
  a_s(\bm{k}_e).
\end{equation}
The corresponding representation for electrons in
periodic potentials is
\begin{eqnarray}\nonumber
  \Psi_s(\bm{x})&=&\sqrt{\frac{V}{8\pi^3}}\sum_n\int_{\mathcal{B}}\!d^3\bm{k}_e\,
  \exp(\mathrm{i}\bm{k}_e\cdot\bm{x})
  \\ \label{eq:Psi2aband}
  &&\times u_n(\bm{k}_e,\bm{x})a_{n,s}(\bm{k}_e),
\end{eqnarray}
where $n$ labels the energy bands, $V$ is the volume of the Wigner-Seitz cell,
$u_n(\bm{k}_e,\bm{x})$ is the Bloch factor with normalization (\ref{eq:orthou}),
and the integration is over the Brillouin zone $\mathcal{B}$.

However, in both cases we are still dealing with electrons which couple to photons
through the quantum optics interaction (\ref{eq:Hegamma}). The inversion of
(\ref{eq:Psi2aband}),
\begin{eqnarray}\nonumber
  a_{n,s}(\bm{k}_e)&=&\sqrt{\frac{V}{8\pi^3}}\int\!d^3\bm{x}\,
  \exp(-\,\mathrm{i}\bm{k}_e\cdot\bm{x})
  \\
  &&\times u_n^+(\bm{k}_e,\bm{x})\Psi_s(\bm{x}),
\end{eqnarray}
provides the unitary transformation from the free electron operators in $\bm{k}_e$-space
to the operators $a_{n,s}(\bm{k}_e)$ for electrons in energy bands,
\begin{eqnarray}\nonumber
  a_{n,s}(\bm{k}_e)&=&\frac{\sqrt{V}}{8\pi^3}\int\!d^3\bm{x}\int\!d^3\bm{k}'_e\,
  \exp[\mathrm{i}(\bm{k}'_e-\bm{k}_e)\cdot\bm{x}]
  \\
  &&\times u_n^+(\bm{k}_e,\bm{x})a_s(\bm{k}'_e).
\end{eqnarray}
The different $\bm{k}_e$-space operators correspond to different convenient
representations of electron operators depending on whether the electrons are moving freely
or in a periodic potential.

With respect to the momenta of electrons moving in a periodic potential, we note that
the equation for free electrons,
$\langle\bm{k}'_e|\op{p}|\bm{k}_e\rangle=\hbar\bm{k}_e\delta(\bm{k}_e-\bm{k}'_e)$
gets modified to
\begin{eqnarray} \nonumber
  &&\langle n,\bm{k}'_e|\op{p}|n,\bm{k}_e\rangle=\delta(\bm{k}_e-\bm{k}'_e)
  \\
  &&\quad\times\!\left(\hbar\bm{k}_e
  +\int_V\!d^3\bm{x}\, u^+_n(\bm{k}_e,\bm{x})
  \frac{\hbar}{\mathrm{i}}\frac{\partial}{\partial\bm{x}}u_n(\bm{k}_e,\bm{x})\right)\!,
\end{eqnarray}
where the remaining integration on the right hand side is over the Wigner-Seitz cell.
The limitation of translation symmetry to lattice vectors due to
the periodic lattice potential implies that the energy eigenstates are not momentum
eigenstates anymore. However, $\op{p}$ is still the first-quantized momentum operator for
the electrons in the Bloch energy eigenstates.

We gave the Schr\"odinger picture electron states with their time-dependence
in Eqs.~(\ref{eq:1e},\ref{eq:Ne}). However,
the states only enter for time $t=0$ into the transition matix elements, because the unperturbed
time evolution operators $U_0(t,0)$ from the external states
transform the full time evolution operator in the
scattering matrix elements into the time evolution operator of the interaction picture,
evaluated between states at time $t=0$.

Note that the bold-face kets in (\ref{eq:1e},\ref{eq:Ne}) correspond to Fock space states,
which appear in the derivations of the transition amplitudes reported in the body of this review, but
not in the final results. The kets in the transition amplitudes in Secs.~\ref{sec:em}-\ref{sec:sc}
correspond to the wave functions of the first quantized theory that enter into the matrix elements,
e.g.~$\psi_s(\bm{x})=\langle\bm{x}|\psi_s\rangle$.

\section{Atomic recoil}

Recoils are usually neglected in transition matrix elements involving discrete
atomic states. Recoil effects will be more prominent for transitions involving free
atoms than for atoms which are bound into a molecule or a lattice. We will therefore
discuss recoils between states of free atoms.

In a description
of single electrons moving in an effective potential $V(\bm{x}_e-\bm{x}_p)$
created by the nucleus at position $\bm{x}_p$ (the notation is motivated by
exactness of these considerations for hydrogen atoms) and a radially symmetric
distribution of the other electrons, the eigenstates can be written in the form
\begin{eqnarray}\nonumber
  \langle\bm{x}_e,\bm{x}_p|\bm{K},n\rangle&=&\frac{1}{\sqrt{2\pi}^3}
  \exp\!\left(\mathrm{i}\bm{K}\cdot\frac{m_p\bm{x}_p+m_e\bm{x}_e}{m_p+m_e}\right)
  \\ \label{eq:2pstates}
  &&\times
  \psi_n(\bm{x}_e-\bm{x}_p).
\end{eqnarray}
Here $\hbar\bm{K}$ is the center-of-mass momentum
and $n$ represents the remaining set of orbital and spin quantum numbers
of this effective two-particle
problem. The mass $m_p\gg m_e$ includes the mass of the nucleus.
The energy of the state (\ref{eq:2pstates}) is
\begin{equation}
E_n(\bm{K})=\frac{\hbar^2\bm{K}^2}{2(m_p+m_e)}+\hbar\omega_n.
\end{equation}

Scattering matrix
elements for transitions $|\bm{K}_i,n_i\rangle\to |\bm{K}_f,n_f\rangle$ due
to electron-photon coupling then involve transition matrix elements
\begin{eqnarray}\nonumber
  &&\langle\bm{K}_f,n_f|\exp(-\,\mathrm{i}\bm{k}\cdot\op{x}_e)
  \bm{\epsilon}^+_\gamma(\bm{k})\cdot\frac{\op{p}_e}{m_e}
  |\bm{K}_i,n_i\rangle
  \\ \nonumber
&&=
  \int\!\frac{d^3\bm{x}_p}{(2\pi)^3}\int\!d^3\bm{x}_e\,
  \exp\!\left(\mathrm{i}(\bm{K}_i-\bm{K}_f)\cdot\frac{m_p\bm{x}_p+m_e\bm{x}_e}{m_p+m_e}\right)
  \\ \nonumber
  &&\quad\times
  \psi^+_f(\bm{x}_e-\bm{x}_p)\exp(-\,\mathrm{i}\bm{k}\cdot\bm{x}_e)
  \\ 
  && \quad\times\bm{\epsilon}^+_\gamma(\bm{k})\cdot\left(\frac{\hbar}{\mathrm{i}m_e}
  \frac{\partial}{\partial\bm{x}_e}+\frac{m_e\hbar\bm{K}_i}{m_p+m_e}\right)
  \psi_i(\bm{x}_e-\bm{x}_p).
\end{eqnarray}

With regard to the photon terms, this is formulated for the case of emission
of a real or virtual photon through this matrix element, but the argument
of negligibility of atomic recoils given below is equally valid for
the substitutions which correspond to photon
absorption, $\bm{\epsilon}^+_\gamma(\bm{k})\to\bm{\epsilon}_\gamma(\bm{k})$,  
$\exp(-\,\mathrm{i}\bm{k}\cdot\op{x}_e)\to\exp(\mathrm{i}\bm{k}\cdot\op{x}_e)$.

In the next step, we substitute center of mass and relative coordinates
\begin{equation}
  \bm{X}=\frac{m_p\bm{x}_p+m_e\bm{x}_e}{m_p+m_e},\quad
  \bm{x}=\bm{x}_e-\bm{x}_p,
\end{equation}
\begin{equation}
\bm{x}_e=\bm{X}+\frac{m_p}{m_p+m_e}\bm{x},
  \end{equation}
\begin{equation}
\frac{\partial}{\partial\bm{x}_e}
=
\frac{\partial}{\partial\bm{x}}+\frac{m_e}{m_p+m_e}\frac{\partial}{\partial\bm{X}},
\end{equation}
use dipole approximation in the atomic state matrix element,
\begin{equation}
\exp\!\left(-\,\mathrm{i}\bm{k}\cdot\frac{m_p}{m_p+m_e}\bm{x}\right)\to 1,
\end{equation}
and use orthogonality of the initial and final
states, $\langle\bm{K}_f,n_f|\bm{K}_i,n_i\rangle=0$,
to find in dipole approximation
 \begin{eqnarray}\nonumber
  &&\langle\bm{K}_f,n_f|\exp(-\,\mathrm{i}\bm{k}\cdot\op{x}_e)
  \bm{\epsilon}^+_\gamma(\bm{k})\cdot\frac{\op{p}_e}{m_e}
  |\bm{K}_i,n_i\rangle 
  \\ \nonumber
  &&=\frac{1}{(2\pi)^3}\int\!d^3\bm{X}\,
  \exp[\mathrm{i}(\bm{K}_i-\bm{K}_f-\bm{k})\cdot\bm{X}]
  \\  \nonumber
  &&\quad\times\int\!d^3\bm{x}\,
  \psi^+_f(\bm{x})\bm{\epsilon}^+_\gamma(\bm{k})\cdot
  \frac{\hbar}{\mathrm{i}m_e}\frac{\partial}{\partial\bm{x}}\psi_i(\bm{x})
  \\ \nonumber
  &&=\delta(\bm{K}_i-\bm{K}_f-\bm{k})
  \\ 
  &&\quad\times\int\!d^3\bm{x}\,
  \psi^+_f(\bm{x})\bm{\epsilon}^+_\gamma(\bm{k})\cdot
  \frac{\hbar}{\mathrm{i}m_e}\frac{\partial}{\partial\bm{x}}\psi_i(\bm{x}).
 \end{eqnarray}

 Inclusion of the center of mass motion of the atom multiplies the standard
 transition matrix element with a momentum conserving $\delta$ function.
 The squares of these extra factors cancel in the calculation of observables
 from the squares $|S_{fi}|^2$ of scattering matrix elements, since the final
 center of mass momentum $\hbar\bm{K}_f$ adds an integration $\int d^3\bm{K}_f\ldots$,
 the square of the momentum conserving $\delta$ function contributes a factor
 $(\mathcal{V}/8\pi^3)\delta(\bm{K}_i-\bm{K}_f-\bm{k})$, and the fixed initial
 center of mass momentum contributes an elementary $\bm{K}$ space
 volume $8\pi^3/\mathcal{V}$,
 such that the only contribution from inclusion of the center of mass motion
 is to shift $\bm{K}_i\to\bm{K}_f=\bm{K}_i-\bm{k}$. This shifts energy
 conservation for spontaneous emission, $E_i(\bm{K}_i)>E_f(\bm{K}_f)$, from
 $ck=\omega_{if}=[E_i(\bm{K})-E_f(\bm{K})]/\hbar$ to
 \begin{eqnarray}\nonumber
   ck&=&\frac{E_i(\bm{K}_i)-E_f(\bm{K}_f)}{\hbar}
   \\
   &=&\frac{\hbar}{2}\frac{2\bm{K}_i\cdot\bm{k}-\bm{k}^2}{m_p+m_e}
   +\omega_i-\omega_f.
 \end{eqnarray}
 The net effect of the atomic recoil is therefore
 a frequency shift that in leading order in $\hbar|\bm{K}_i|/(m_p+m_e)c$
 and $\hbar\omega_{if}/(m_p+m_e)c^2$
 takes the form (with scattering angle $\theta$)
 \begin{equation}\label{eq:DeltaE}
   ck=\omega_{if}+\frac{\hbar|\bm{K}_i|\cos\theta}{(m_p+m_e)c}\omega_{if}
   -\frac{\hbar\omega_{if}^2}{2(m_p+m_e)c^2}.
 \end{equation}
The frequency shift is very small in comparison to $\omega_{if}$ if the transition involves 
nonrelativistic atoms, $\hbar|\bm{K}_i|\ll m_pc$. In cases of absorption or scattering,
when we have photons in the initial state, we should also exclude high
energy $\gamma$-rays to neglect atomic recoil effects. However, recall that we have already
excluded photons beyond the soft X-ray regime through the use of dipole approximation.
We also note that for photons below the hard X-ray regime, possible energy shifts from
recoils of quasifree electrons are also suppressed by $\hbar|\bm{k}|/m_ec$.

The observation that recoils yield frequency shifts, but do not generate
extra factors in transition rates, also applies if the matrix element
 does not directly connect initial and final atomic states but
 includes intermediate virtual states.
 Integration over intermediate virtual center of mass momenta $\hbar\bm{K}_v$
 only reduces
 momentum conserving $\delta$ functions in $|S_{fi}|^2$ according to
 \begin{eqnarray}\nonumber
   &&\int\!d^3\bm{K}_v\,\delta(\bm{K}_v-\bm{K}_f-\bm{k}')\delta(\bm{K}_i-\bm{K}_v+\bm{k})
   \\ 
   &&=\delta(\bm{K}_i-\bm{K}_f+\bm{k}-\bm{k}').
   \end{eqnarray}

\section{Photon emission from radiative electron-hole recombination between energy bands}
\label{sec:eh}

Eqs.~(\ref{eq:ecck},\ref{eq:ecck2}) apply to transitions
between conduction bands, but not to a conduction electron filling a valence band hole.
Electron-hole annihilation through interband transition requires special considerations
because now we are effectively dealing with two particles in the initial 
state, \textit{viz.}~a conduction electron with wave vector $\bm{k}_{e,i}$ and a
valence band hole 
with wave vector $-\,\bm{k}_{e,f}=\Delta\bm{k}-\bm{k}_{e,i}$.
Interband electron-hole recombination is therefore akin to two-particle annihilation
events that are characterized by annihilation cross sections.

We  use subscripts $c$
and $v$ for labeling the conduction band and the valence band, respectively.  
The initial electron-hole state
is then $a^+_{c,s}(\bm{k}_{e,i})a_{v,s}(\bm{k}_{e,f})\bm{|}\Omega\bm{\rangle}$,
where the ground state $\bm{|}\Omega\bm{\rangle}$ corresponds to the filled Fermi volume
in the Brillouin zone, $a^+_{c,s}(\bm{k}_{e,i})$ is an electron creation operator in the
conduction band, and $c^+_{v,s}(-\,\bm{k}_{e,f})=a_{v,s}(\bm{k}_{e,f})$ acts as a hole
creation operator in the valence band \cite{kittel,holenote}.
We also assume the same spin projection $s$ for the electron and the hole
because the operators (\ref{eq:Hegamma}) preserve spin projections.

The reduction in final state measure due to the initial
state $a^+_{c,s}(\bm{k}_{e,i})a_{v,s}(\bm{k}_{e,f})\bm{|}\Omega\bm{\rangle}$, compared to the
case in Sec.~\ref{sec:ecc} of transition between conduction
bands, $d^3\bm{k} d^3\bm{k}_{e,f}\to d^3\bm{k}$,
implies that integration over final states now only
takes care of momentum conservation,
but an energy-conserving $\delta$-function remains,
\begin{eqnarray}\nonumber
&&\delta[ck+\omega_v(\bm{k}_{e,f})-\omega_c(\bm{k}_{e,i})]
  \delta(\bm{k}+\bm{k}_{e,f}-\bm{k}_{e,i})
  \\ \label{eq:ehenergy}
  &&\to\delta[c|\bm{k}_{e,i}-\bm{k}_{e,f}|+\omega_v(\bm{k}_{e,f})-\omega_c(\bm{k}_{e,i})],
\end{eqnarray}
and this constrains the pairs of wave vectors in the Brillouin zone where
electron-hole recombination generates
single-photon emission as a purely radiative process.
Indeed, we should expect kinematic constraints since the corresponding
process for free particle-antiparticle pairs is forbidden by
energy-momentum conservation.
Single-photon emission from electron-hole recombination for generic
combinations of wave vectors therefore requires
phonon assistance, creation of Auger electrons, assistance
through particle traps, or two-photon emission.
These mechanisms are outside of the scope
of the current review of leading order radiative processes.

 For completeness, however, we give a formula that applies to the purely
radiative single-photon emission from electron-hole recombination if the
constraint (\ref{eq:ehenergy}) can be satisfied within the linewidth of the
transition.

Electron-hole recombination rates are normalized by the differential current
density of conduction electrons
\begin{equation}
\frac{d\bm{j}_e(\bm{k}_e)}{d^3\bm{k}_e}=\frac{1}{V}\int_V\!d^3\bm{x}\,
\frac{d\bm{{J}}_e(\bm{k}_e,\bm{x})}{d^3\bm{k}_e},
\end{equation}
which arises from the $\bm{x}$-dependent differential current density of conduction
electrons,
\begin{eqnarray}\nonumber
  d\bm{{J}}_e(\bm{k}_e,\bm{x})&=&\psi_c^+(\bm{k}_e,\bm{x})\frac{\hbar}{2\mathrm{i}m}
  \frac{\partial\psi_c(\bm{k}_e,\bm{x})}{\partial\bm{x}}d^3\bm{k}_e
\\
&&-\,\frac{\hbar}{2\mathrm{i}m}\frac{\partial\psi_c^+(\bm{k}_e,\bm{x})}{\partial\bm{x}}
\psi_c(\bm{k}_e,\bm{x})d^3\bm{k}_e,
\end{eqnarray}
through averaging over the Wigner-Seitz cell. Note that the differential current densities
$d\bm{{J}}_e(\bm{k}_e,\bm{x})/d^3\bm{k}_e$ and $d\bm{j}_e(\bm{k}_e)/d^3\bm{k}_e$ have
dimensions of velocities, e.g.~units of cm/s.

Radiative electron-hole recombination through single-photon emission
can then be characterized by a cross section
\begin{eqnarray}\nonumber
  &&\sigma(\bm{k}_{e,i},\bm{k}_{e,f},\bm{\epsilon}_\gamma)=
  \frac{\alpha_S}{2\pi}\frac{[\omega_c(\bm{k}_{e,i})-\omega_v(\bm{k}_{e,f})]^2}{
    |\bm{k}_{e,i}-\bm{k}_{e,f}|}
  \\ \nonumber
  &&\quad\times\frac{|\langle u_v(\bm{k}_{e,f})|\bm{\epsilon}^+_\gamma(\bm{k}_{e,i}-\bm{k}_{e,f})
    \cdot\op{x}|u_c(\bm{k}_{e,i})\rangle_V|^2}{|d\bm{j}_e(\bm{k}_{e,i})/d^3\bm{k}_{e,i}|}
  \\
  &&\quad\times\Delta_\Gamma[c|\bm{k}_{e,i}-\bm{k}_{e,f}|+\omega_v(\bm{k}_{e,f})-\omega_c(\bm{k}_{e,i})].
\end{eqnarray}

As explained in Eqs.~(\ref{eq:total1})
and (\ref{eq:avproof1}-\ref{eq:avproof4}), averaging over angles removes
dependence of matrix elements on polarization vectors. Here this yields an angle averaged
electron-hole recombination cross section from single-photon emission,
\begin{eqnarray}\nonumber
  &&\sigma(\bm{k}_{e,i},\bm{k}_{e,f})=
  \frac{\alpha_S}{6\pi}\frac{[\omega_c(\bm{k}_{e,i})-\omega_v(\bm{k}_{e,f})]^2}{
    |\bm{k}_{e,i}-\bm{k}_{e,f}|}
  \\ \nonumber
  &&\quad\times\frac{|\langle u_v(\bm{k}_{e,f})|
    \op{x}|u_c(\bm{k}_{e,i})\rangle_V|^2}{|d\bm{j}_e(\bm{k}_{e,i})/d^3\bm{k}_{e,i}|}
  \\
  &&\quad\times\Delta_\Gamma[c|\bm{k}_{e,i}-\bm{k}_{e,f}|+\omega_v(\bm{k}_{e,f})-\omega_c(\bm{k}_{e,i})].
\end{eqnarray}

Integration of $\sigma(\bm{k}_{e,i},\bm{k}_{e,f})$
against the $\bm{k}_{e,i}$-dependent differential
conduction current density $|d\bm{j}_e(\bm{k}_{e,i})/d^3\bm{k}_{e,i}|$
and the $\bm{k}_{e,f}$-dependent differential hole density
$|d\rho_h(\bm{k}_{e,f})/d^3\bm{k}_{e,f}|$ (which is $\bm{x}$-independent
if averaged over the Wigner-Seitz cell) then yields an estimate for the photon emission
rate per volume for those photons that were generated through purely radiative
single-photon electron-hole recombination, i.e.~from recombinations that were not
assisted through phonon processes or Auger excitations or trapping mechanisms,
and that did not result from two-photon emission.

\acknowledgments
We would like to thank Alexander Moewes, Graham George and Robert Green for encouraging
and helpful comments. 
We acknowledge support from the Natural Sciences and Engineering Research Council
of Canada.\\



\begin{thebibliography}{88}
\bibitem{notequasi1}
{Excitations within an energy band and near an extremum could eventually
  be described through quasiparticles with an effective mass $m_\ast\neq m$. The advantage of such a
  quasiparticle description of intraband excitations near an extremum is replacement of the
  actual energy band $E_n(\bm{k}_e)$ with a parabolic
  band $E_\ast(\bm{k}_e)=E_\ast(\bm{0})+\hbar^2\bm{k}^2_e/2m_\ast$,
  and replacement of the electronic Bloch energy eigenfunctions $\psi_n(\bm{k}_e,\bm{x})$ with 
  plane waves. However, this is not
possible for interband transitions that are not confined to extrema with identical effective masses.
Please also note that Bloch energy eigenfunctions and energy bands in materials are derived from
Hamiltonians (\ref{eq:Hfirstq}) with $m=m_e$, not from Hamiltonians containing effective masses.
Stated differently, electrons in energy bands in materials are still electrons, not quasielectrons.
The situation is not different from electrons in other complex many-electron systems like
many-electron atoms or molecules.}
\bibitem{holenote}
  {However, as pointed out already in note \cite{notequasi1}, we cannot in general
  use a hole quasifermion picture for interband transitions, but still have
  to use the electronic Bloch wave function for the missing electron.}
\end{thebibliography}
\end{document}